\documentclass[12pt,preprint]{aastex}

\newcommand{\mone}{$^{-1}$}
\newcommand{\mtwo}{$^{-2}$}

\newcommand\asec{^{\prime\prime}}

\newcommand\lae{\mathrel{<\kern-1.0em\lower0.9ex\hbox{$\sim$}}}
\newcommand\gae{\mathrel{>\kern-1.0em\lower0.9ex\hbox{$\sim$}}}

\begin{document}
\title{VLBA and Chandra Observations of Jets in FRI radio galaxies: Constraints on Jet
Evolution}

\author{P. Kharb}
\affil{Department of Physics, Rochester Institute of Technology (R.I.T.), 54 Lomb Memorial Drive, Rochester, NY 14623}
\author{C. P. O'Dea}
\affil{Department of Physics, R.I.T., 54 Lomb Memorial Drive, Rochester, NY 14623}
\affil{Harvard-Smithsonian Center for Astrophysics, 60 Garden Street, Cambridge, MA 02138}
\author{A. Tilak}
\affil{Research Associate, Smithsonian Astrophysical Observatory, 60 Garden Street, Cambridge, MA 02138}
\author{S. A. Baum}
\affil{Center for Imaging Science, R.I.T., 54 Lomb Memorial Drive, Rochester, NY 14623 Radcliffe Institute for Advanced Study, 10 Garden Street, Cambridge, MA 02138}
\author{E. Haynes}
\affil{Center for Imaging Science, R.I.T., Rochester, NY 14623}
\author{J. Noel-Storr}
\affil{Center for Imaging Science, R.I.T., 54 Lomb Memorial Drive, Rochester, NY 14623}
\author{C. Fallon}
\affil{Center for Imaging Science, R.I.T., Rochester, NY 14623}
\author{K. Christiansen}
\affil{Center for Imaging Science, R.I.T., Rochester, NY 14623}

\begin{abstract}
We present here the results from new Very Long Baseline Array observations at 1.6 and 5\,GHz of 19 galaxies of a complete sample of 21 UGC FRI radio galaxies. New {\it Chandra} data of two sources, {\it viz.}, UGC\,00408 and UGC\,08433, are combined with the {\it Chandra} archival data of 13 sources. The 5\,GHz observations of ten ``core-jet" sources are polarization-sensitive, while the 1.6\,GHz observations constitute second epoch total intensity observations of nine ``core-only" sources. 
Polarized emission is detected in the jets of seven sources at 5\,GHz, but the cores are essentially unpolarized, {except in M\,87}. Polarization is detected at the jet edges in several sources, and the inferred magnetic field is primarily aligned with the jet direction. This could be indicative of magnetic field ``shearing'' due to jet-medium interaction, or the presence of helical magnetic fields. The jet peak intensity $I_\nu$ falls with distance $d$ from the core, following the relation, $I_\nu\propto d^{a}$, where $a$ is typically $\sim-1.5$. Assuming that adiabatic expansion losses are primarily responsible for the jet intensity ``dimming'', two limiting cases are considered: [1] the jet has a constant speed on parsec-scales and is expanding gradually such that the jet radius $r\propto d^{0.4}$; this expansion is however unobservable in the laterally unresolved jets at 5\,GHz, and [2] the jet is cylindrical and is accelerating on parsec-scales. Accelerating parsec-scale jets are consistent with the phenomenon of ``magnetic driving'' in Poynting flux dominated jets. {While slow jet expansion as predicted by case [2] is indeed observed in a few sources from the literature that are resolved laterally, on scales of tens or hundreds of parsec, case [2] cannot be ruled out in the present data, provided the jets become conical on scales larger than those probed by the VLBA.}
{\it Chandra} observations of 15 UGC FRIs detect X-ray jets in nine of them. The high frequency of occurance of X-ray jets in this complete sample suggests that they are a signature of a ubiquitous process in FRI jets. It appears that the FRI jets start out relativistically on parsec-scales but decelerate on kiloparsec scales, with the X-ray emission revealing the sites of bulk deceleration and particle reacceleration.
\end{abstract}

\section{Introduction}
It is now widely believed that the enormous energy output from an active galactic nucleus (AGN) is the result of mass accretion onto a supermassive black hole. A small fraction of AGNs have powerful bipolar radio outflows extending to intergalactic distances. These ``radio-loud'' AGN fall primarily into two morphological classes: the relatively low power Fanaroff-Riley type I \citep[FRI;][]{Fanaroff74} radio galaxies possess broad jets that flare out into radio lobes, while the high power FRII radio galaxies exhibit narrow, collimated jets that terminate in hot spots, with the backflowing plasma forming the radio lobes. As AGN jets experience bulk relativistic motion, which gives rise to Doppler favoritism, and most have axisymmetric tori-like obscuring structures (though perhaps not all, e.g., FRI galaxies - \cite{chia99,Kharb04A}), which hide their central regions from certain lines of sight, their orientation in the sky plays a dominant role in their appearance. On the basis of orientation-independent properties like extended radio emission, emission line spectra, host galaxy type and galaxy environment, a simple Unified Scheme has emerged \citep[e.g.,][]{UrryPadovani95}, according to which the BL~Lac objects and radio-loud quasars are the relativistically beamed counterparts of FRI and FRII radio galaxies, respectively \citep[however see][]{Singal96,Kharb10}. 

It has been suggested that the jets in both classes start out relativistically \citep[e.g.,][]{gio01}, but the FRI jets decelerate on scales of about a kiloparsec from the core \citep[e.g.,][]{Laing99}. The aim of this paper is to explore the scales at which FRI jets decelerate through parsec-scale total and polarimetric observations with the Very Long Baseline Array (VLBA) and kiloparsec-scale X-ray observations with the {\it Chandra} X-ray Observatory.

\subsection{UGC Sample and Previous Work}
In order to investigate the nuclear environment and jet physics of low power radio galaxies, we have been conducting a large multi-wavelength study of a well-defined complete sample of 21 FRI radio galaxies from the Uppasala General Catalog of galaxies (Table~\ref{sample}).
The selection criteria are:
(i) Hubble type E or S0;
(ii) recession velocity $<$7000 km\,s$^{-1}$ ($z<0.0229$);
(iii) optical major axis diameter $>1\arcmin$;
(iv) total flux density at 1.4 GHz, $S_{1.4}>$150 mJy;
(v) declination $-5\degr<\delta<82\degr$;
(vi) black hole rather than starburst source from the IRAS\footnote{Infrared Astronomical Satellite}/radio flux ratio; 
(vii) radio size $>10\arcsec$ in Very Large Array (VLA) images at 1.49\,GHz with 2$\arcsec$ resolution, where the size is measured at contours of 3$\sigma$ \citep{cond88}.
The sample covers two magnitudes in absolute blue luminosity and about 80 in radio power. 

The host galaxies have been studied by \cite{ver99,ver02} using the Wide-Field Planetary Camera 2 (WFPC2) aboard the {\it Hubble Space Telescope} ({HST}), while the gas dynamics and black hole masses have been studied by \cite{noe03} using the Space Telescope Imaging Spectrograph (STIS) on {HST}. As the selection criteria specify a constraint on the total radio luminosity, which is typically dominated by the kiloparsec scale lobe structure, the sample is unbiased in radio core luminosity. 

The 1.6\,GHz VLBA$-$VLA study of the radio structure in 17 UGC galaxies was carried out by \cite{xu00}. The four sources that were not included in the study were the well-studied source UCG\,07654 (M\,87), UGC\,6635 which did not have a VLA core, and UGC\,07115 and UGC\,12064, which were included in the sample after the VLBA observations had been scheduled. Parsec-scale radio emission was detected in all 17 sources. Of these, five sources showed only cores, 10 showed ``core-jet'' structures, and two sources showed ``twin-jet'' structures. 
\citet{xu00} found that the VLBA jets were aligned with the VLA jets. For the ``twin-jet'' sources, the brighter VLBA jet pointed towards the brighter VLA jet. Furthermore, the jet to counterjet surface brightness ratio, measured in the VLBA data was greater than in the VLA data. All the above properties (including the statistics on the source structures) indicated Doppler boosting on parsec-scales and deceleration on kiloparsec-scales. The observations also revealed that the parsec-scale jet luminosity per unit length decreased or ``dimmed" with distance $r$, as $L_{jet}/pc\sim r^{-2}$. Expansion losses in an adiabatically expanding jet with constant velocity and opening angle, with a magnetic field that was primarily perpendicular to the jet direction was favored to be the jet ``dimming" mechanism.

In order to further investigate these issues and look for the magnetic field configuration in these sources, we obtained polarization-sensitive VLBA observations of ten ``core-jet'' sources at 5\,GHz (excluding the two well-studied sources with jets, {\it viz.}, UGC\,00597  (NGC\,315), and UGC\,07360 (3C\,270)). In order to search for possible weak jet emission, we reobserved nine mostly ``core-only'' sources at 1.6\,GHz with the VLBA. These nine sources included the two sources that were included later on in the sample, UGC\,06635 which lacked a VLA core, and UGC\,12531 which was only tentatively classified as a ``core-jet'' source by \citet{xu00}. 

15 sample UGC galaxies have also been studied with the {\it Chandra} X-ray Observatory. Nine of these 15 sources exhibit X-ray jets in them. We present here new {\it Chandra} X-ray images of two sample FRI radio galaxies, {\it viz.,} UGC\,00408 and UGC\,08433.

The paper is structured as follows. \S2 and \S3 describe the radio and X-ray observations, data-analysis and results, respectively. The discussion follows in \S4, while \S5 presents the summary and conclusions from this study. The spectral index $\alpha$ is defined such that flux density at frequency $\nu$ is, $S_{\nu}\propto\nu^{-\alpha}$ and the photon index $\Gamma=1+\alpha$. Throughout this paper, we adopt the cosmology in which $H_0$=71 km s$^{-1}$ Mpc$^{-1}$, $\Omega_m$=0.27 and $\Omega_{\Lambda}$=0.73.

\begin{deluxetable}{ccccccc}
\tabletypesize{\scriptsize}
\tablecaption{UGC sample}
\tablewidth{0pt}
\tablehead{
\colhead{Source} &\colhead{Other} &\colhead{Galaxy} & \colhead{R.A.}& \colhead{Dec}& \colhead{Redshift} & \colhead{Scale}\\
\colhead{Name}&\colhead{Names}&\colhead{Type}&\colhead{hr m s}&\colhead{$\degr~\arcmin~\arcsec$}&\colhead{}&\colhead{pc/mas}}
\startdata
UGC\,00408& NGC\,193& S0& 00 39 18.5829& +03 19 52.584& 0.014657 &0.274\\
UGC\,00597& NGC\,315& E& 00 57 48.8834& +30 21 08.812& 0.016485 &0.311\\
UGC\,00689& 3C\,31, NGC\,383& E0& 01 07 24.9593& +32 24 45.237& 0.017005&0.322\\
UGC\,01004& NGC\,541& E& 01 25 44.3078& $-$01 22 46.522& 0.018086&0.343\\
UGC\,01413& NGC\,741& E& 01 56 20.9902& +05 37 44.277& 0.018549&0.353\\
UGC\,01841& 3C\,66B& E& 02 23 11.4073& +42 59 31.403& 0.021258&0.411\\
UGC\,03695& NGC\,2329& E-S0& 07 09 08.0061& +48 36 55.733& 0.019330&0.392\\
UGC\,05073& NGC\,2892& E& 09 32 52.9316& +67 37 02.630& 0.022822&0.460\\
UGC\,06635& NGC\,3801& E& 11 40 17.31& +17 43 36.8& 0.011064&0.246\\
UGC\,06723& 3C\,264, NGC\,3862& E& 11 45 05.0099& +19 36 22.756& 0.021718&0.455\\
UGC\,07115& MCG 04-29-031 & E& 12 08 05.81& +25 14 16.4& 0.022602&0.470\\
UGC\,07360& 3C\,270, NGC\,4261& E& 12 19 23.2162& +05 49 29.702& 0.007465&0.175\\
UGC\,07455& NGC\,4335& E& 12 23 01.8881& +58 26 40.384& 0.015417&0.320\\
UGC\,07494& M\,84, 3C\,272.1, NGC\,4374& S0& 12 25 03.7433& +12 53 13.143& 0.003536&0.095\\
UGC\,07654& M\,87, 3C\,274, NGC\,4486& E& 12 30 49.4234& +12 23 28.044& 0.004360&0.111\\
UGC\,08419& NGC\,5127& E& 13 23 45.0156& +31 33 56.703& 0.015951&0.342\\
UGC\,08433& NGC\,5141& S0& 13 24 51.4403& +36 22 42.763& 0.017379&0.363\\
UGC\,09058& NGC\,5490& E& 14 09 57.2984& +17 32 43.911& 0.016195&0.341\\
UGC\,11718& NGC\,7052& E& 21 18 33.0446& +26 26 49.251& 0.015584&0.293\\
UGC\,12064& 3C\,449& E-S0& 22 31 21.35& +39 21 33.2& 0.017085&0.324\\
UGC\,12531& NGC\,7626& E& 23 20 42.5391& +08 13 00.992& 0.011358&0.205
\enddata
\label{sample}
\end{deluxetable}

\section{VLBA Observations \& Data Reduction} 
Phase-referenced polarization-sensitive observations were carried out for ten ``core-jet'' sources at 4.99\,GHz on 2003, February 7 and 14, with the ten antennas of the VLBA (Project ID: BO015). Phase-referenced VLBA observations for nine other sources were carried out at 1.66\,GHz on 2003, February 5, 8, 11, 15, 16, 20, and 28. A switching cycle of 5 minutes (calibrator 2 minutes and target 3 minutes) was adopted at both the frequencies. The sources were typically observed for 2.0$-$2.5 hours at 5\,GHz, and typically 0.5$-$1.5 hours at 1.6\,GHz. For all the observations, two intermediate frequency (IF) channels with a bandwidth of 8\,MHz each, were used.

The data were reduced and self-calibrated using standard reduction and imaging procedures in the Astronomical Image Processing System (AIPS). The standard steps included fixing the data, applying ionospheric corrections, applying the earth orientation parameters, amplitude calibration and phase calibration. Los Alamos was used as the reference antenna at all stages of the calibration for the 5\,GHz data. In addition, amplitude corrections were made using the AIPS task APCAL while phase corrections for parallel and cross correlated data were made using the task FRING. For the 1.6\,GHz data, the instrumental phase corrections were determined using pulse-cals for all but one dataset {\it viz.,} for the phase reference calibrator J0155+0438 and the target source UGC\,01413. For this one case, we had to use a small subset of the data for fringe fitting to determine the optimal solutions. In addition, Kitt Peak was used as the reference antenna except in three cases where alternate antenae had to be chosen ({\it viz.,} UGC\,01004 $-$ Pie Town; UGC\,01413 and UGC\,06635 $-$ Los Alamos). The instrumental corrections were then applied to the entire dataset using global fringe fitting.

For the 5\,GHz data, the instrumental polarization leakage term calibrator was J1407+2827, while the electric vector position angle (EVPA) calibrator was J1310+3220. The AIPS task LPCAL was used to obtain the antenna D-terms for this calibrator which were then transferred to the main dataset. The EVPA calibrator and its position angle in the VLA monitoring program\footnote{http://www.vla.nrao.edu/astro/calib/polar/} close to the time of the VLBA observation was used to determine the absolute electric vector position angle. The EVPA in the VLA monitoring is assumed to be close to the intrinsic value and the difference between the VLA and the VLBA EVPAs is assumed to be due to instrumental effects which can then be compensated. Specifically, the VLA EVPA information existed for February 8, 2003 and March 5, 2003. Assuming a linear EVPA-Date relation, we obtained the EVPA for the observing dates (February 7 and 14, 2003) via a simple interpolation. The VLBA EVPA was obtained from the polarized regions in the Stokes' Q and U maps of J1310+3220, following the relation $\chi = \frac{1}{2}$tan$^{-1}$(U/Q). The phase reference calibrators were then separated from the main dataset using the task SPLIT. The AIPS tasks IMAGR and CALIB were used iteratively for imaging and self-calibration. The solutions from the final converged images of the phase-calibrators were transferred to the targets. 

Significant discrepancies were observed in the positions of 7 targets at 5\,GHz. The discrepancy was the largest for UGC\,00689 (3C\,31) which was offset by $\sim$120\,mas from its expected position at the center of the map. Incorrect Earth Orientation Parameters (EOP) used in VLBA correlator job scripts\footnote{http://www.vla.nrao.edu/astro/archive/issues/\#503} from early 2003 to 2005, are likely to be responsible for this. At the time of the 5\,GHz data reduction in early 2006, there was no option to correct for this effect in AIPS. For the more recently reduced 1.6\,GHz data however, we could rectify these errors by running the task CLCOR with OPCODE `EOPS'. For the 5\,GHz data, we shifted the targets to the phase center by running the task UVFIX on the SPLIT files, before attempting to create the final self-calibrated images. 

Phase and amplitude self-calibration on the targets improved their images significantly at 5\,GHz. All the 5\,GHz images were created with uniform weighting using a ROBUST parameter = 0 in IMAGR. The Stokes' Q and U maps were used in the task COMB to obtain maps of the polarization intensity and electric vector position angle. Polarization values with S/N$<$3 were blanked in COMB to produce the polarized intensity maps. The polarization angle maps were restricted in COMB to have output errors $<10\degr$. Finally, the fractional polarization maps were created in COMB by constraining the output errors to be $\lesssim10\%$. This latter number was relaxed to 25\% for the fractional polarization image of UGC\,00689. A compilation of the basic observational parameters for each source is given in Table~\ref{vlba_tot}. The integrated flux densities for the cores were obtained in AIPS using the Gaussian-fitting task JMFIT, while the total flux densities were obtained by putting a box around the source, using the AIPS verbs TVWINDOW and IMSTAT. 

The flux density estimates for the second epoch 1.6\,GHz images were obtained from images that were convolved with circular beams of size 10~mas and had UVTAPER set to 15,000 k$\lambda$ in IMAGR. The 1.6\,GHz image of the weakest source, {\it viz.,} UGC\,01004, was created using ROBUST = 5 in IMAGR (signifying natural weighting; the rest of the images had ROBUST = 0). This was done to get estimates in a manner similar to that done by \citet{xu00} for the first epoch 1.6\,GHz observations. Phase self-calibration was not possible (there were too many failed solutions) for these weak targets at 1.6\,GHz, except for UGC\,05073 and UGC\,12531. The final flux density and noise estimates at 1.6\, GHz are tabulated in Table \ref{vlba_sectot}. 

\subsection{Results from the Radio Study}
\begin{figure}
\centering{\includegraphics[width=8cm]{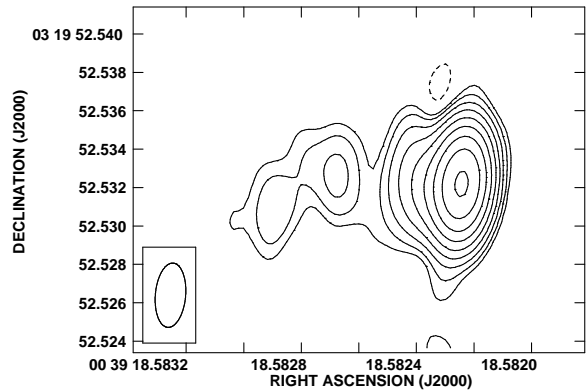}}
\caption[UGC\,00408: VLBA image]{\small 5\,GHz VLBA total intensity image of UGC\,00408. 
Polarization was not detected in this source. Contours are in percentage of peak surface brightness and increase in steps of 2: the lowest contour and peak surface brightness are $\pm$0.35\% and 35\,mJy\,beam$^{-1}$, respectively. }
\label{NGC193_vlba}
\end{figure}

\begin{figure}
\centering{\includegraphics[width=8cm]{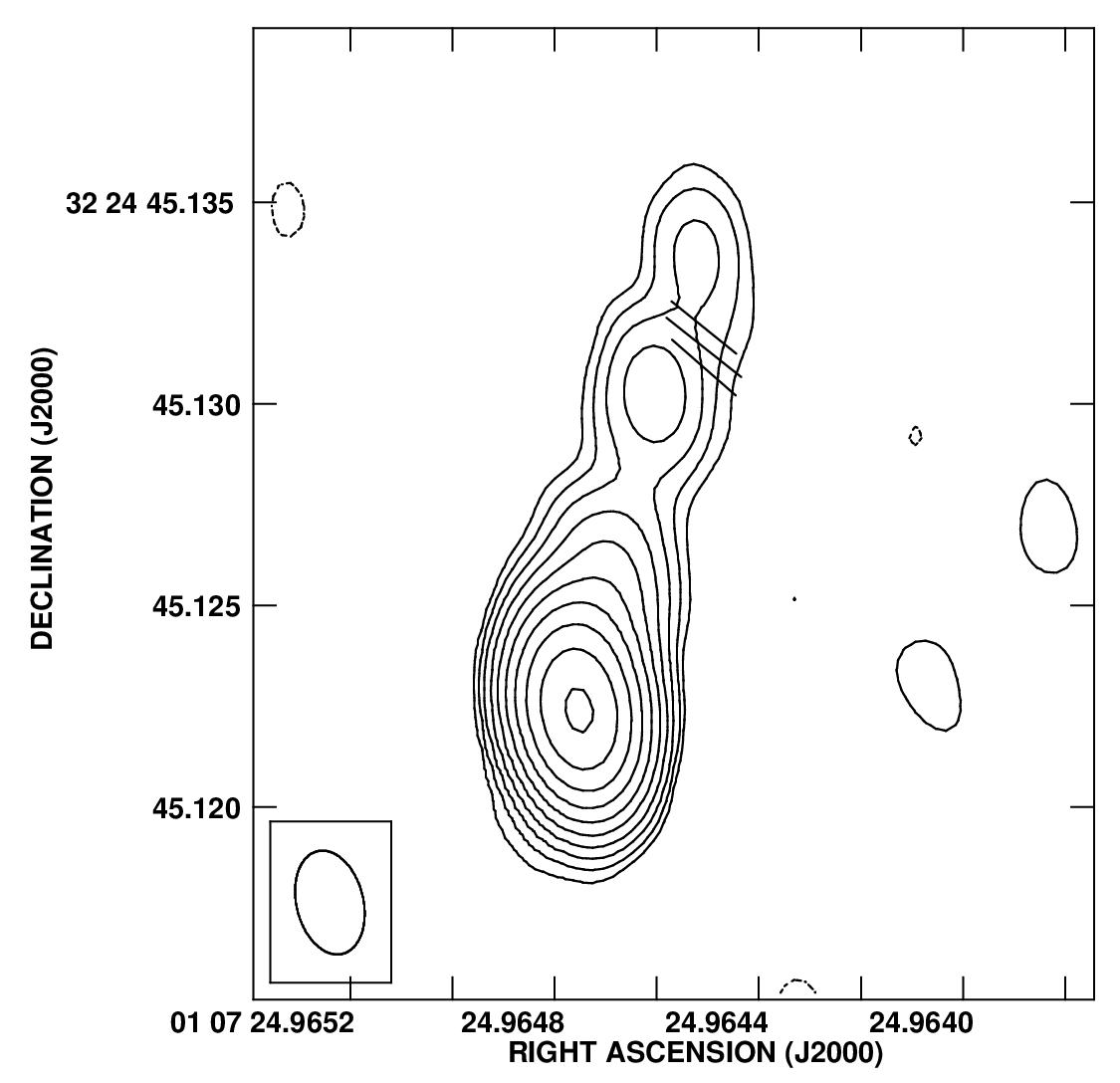}}
\caption[UGC\,00689: VLBA image]{\small 5\,GHz VLBA total intensity image of UGC\,00689 with fractional polarization electric vectors superposed. 1 mas tick = 25\% polarization. Contours are in percentage of peak surface brightness and increase in steps of 2: the lowest contour and peak surface brightness are $\pm$0.17\% and 71\,mJy\,beam$^{-1}$, respectively.}
\label{3C31_vlba}
\end{figure}

\begin{figure}
\centering{\includegraphics[width=8cm]{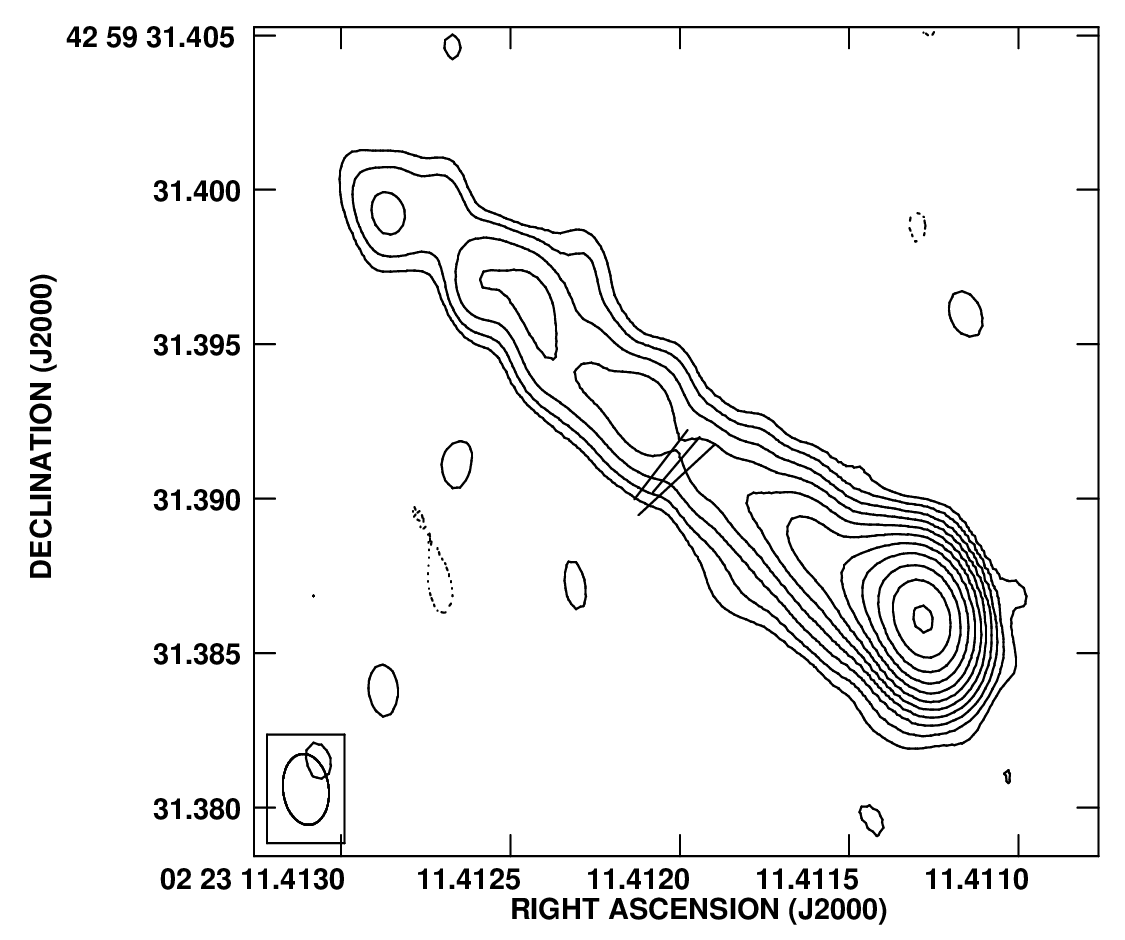}}
\caption[UGC\,01841: VLBA image]{\small 5\,GHz VLBA total intensity image of UGC\,01841 with fractional polarization electric vectors superposed. 1 mas tick = 9\% polarization. Contours are in percentage of peak surface brightness and increase in steps of 2: the lowest contour and peak surface brightness are $\pm$0.085\% and 137\,mJy\,beam$^{-1}$, respectively.}
\label{3C66B_vlba}
\end{figure}

\begin{figure}
\centering{\includegraphics[width=8cm]{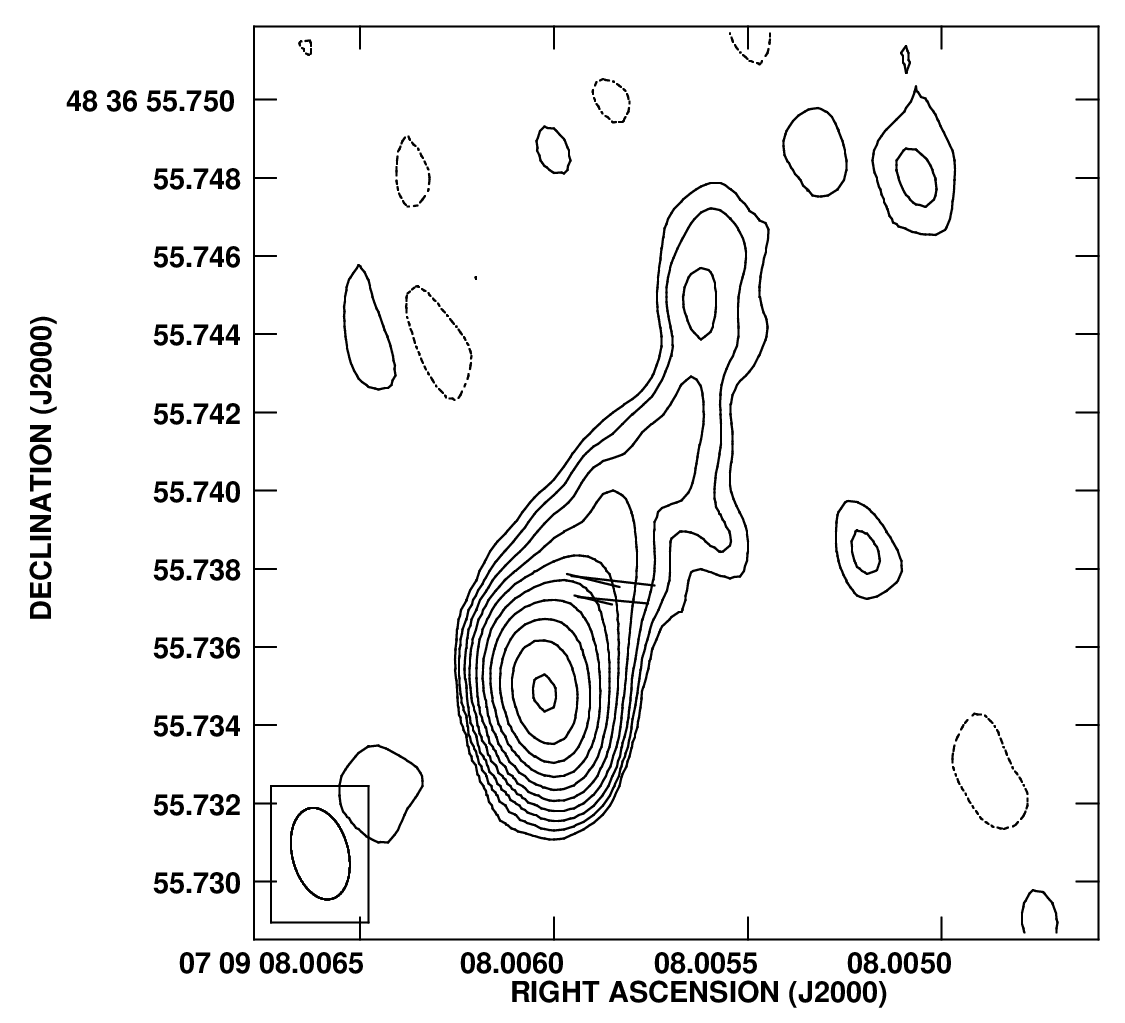}}
\caption[UGC\,03695: VLBA image]{\small 5\,GHz VLBA total intensity image of UGC\,03695 with fractional polarization electric vectors superposed. 1 mas tick = 8\% polarization. Contours are in percentage of peak surface brightness and increase in steps of 2: the lowest contour and peak surface brightness are $\pm$0.35\% and 49\,mJy\,beam$^{-1}$, respectively.}
\label{NGC2329_vlba}
\end{figure}

\begin{figure}
\centering{\includegraphics[width=10cm]{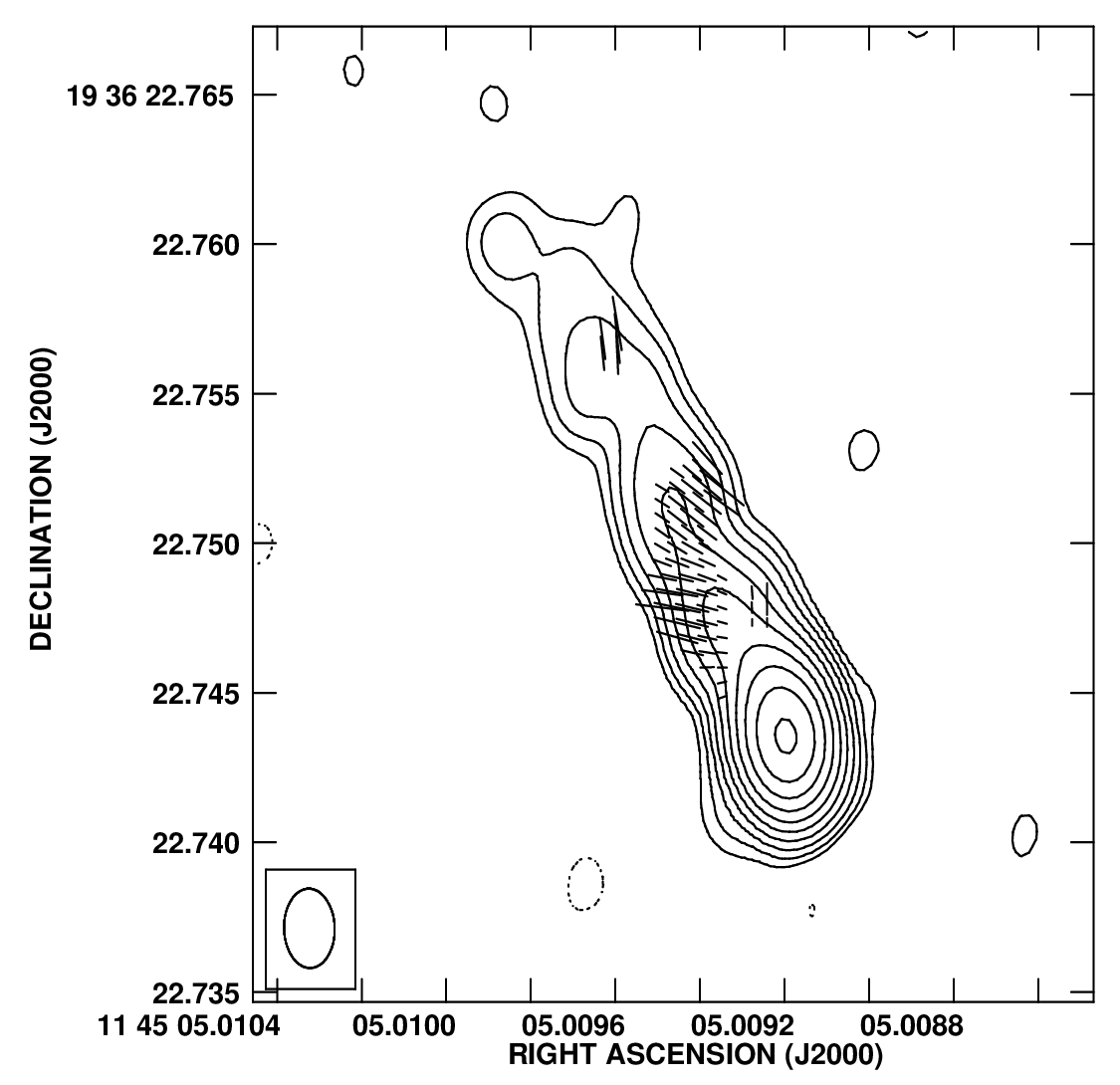}}
\caption[UGC\,06723: VLBA image]{\small 5\,GHz VLBA total intensity image of UGC\,06723 with fractional polarization electric vectors superposed. 1 mas tick = 15\% polarization. Contours are in percentage of peak surface brightness and increase in steps of 2: the lowest contour and peak surface brightness are $\pm$0.17\% and 110\,mJy\,beam$^{-1}$, respectively.}
\label{3C264_vlba}
\end{figure}

\begin{figure}
\centering{\includegraphics[width=10cm]{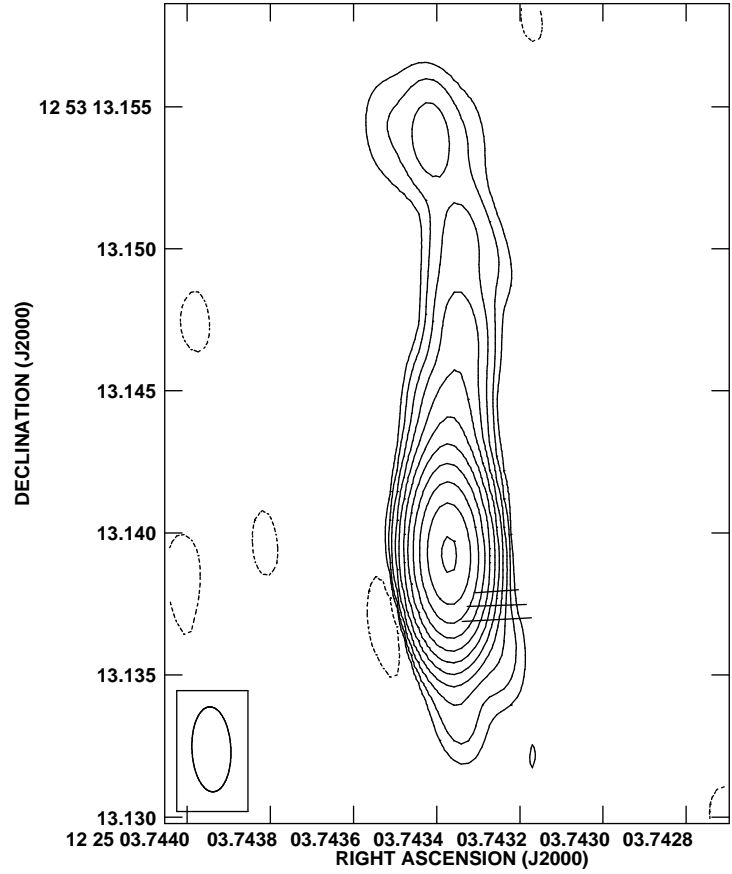}}
\caption[UGC\,07494: VLBA image]{\small 5\,GHz VLBA total intensity image of UGC\,07494 with fractional polarization electric vectors superposed. 1 mas tick = 6\% polarization. Contours are in percentage of peak surface brightness and increase in steps of 2: the lowest contour and peak surface brightness are $\pm$0.085\% and 172\,mJy\,beam$^{-1}$, respectively.}
\label{M84_vlba}
\end{figure}

\begin{figure}
\centering{\includegraphics[width=10cm]{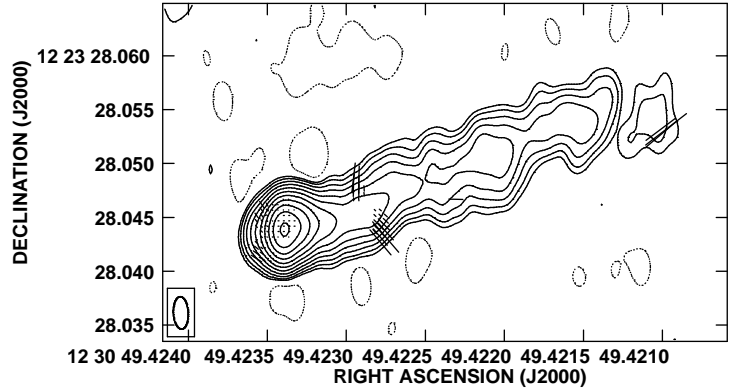}}
\caption[UGC\,07654: VLBA image]{\small 5\,GHz VLBA total intensity image of UGC\,07654 or M\,87 with fractional polarization electric vectors superposed. 1 mas tick = 4\% polarization. Used as phase reference calibrator for UGC\,07494. Contours are in percentage of peak surface brightness and increase in steps of 2: the lowest contour and peak surface brightness are $\pm$0.085\% and 1.06\,Jy\,beam$^{-1}$, respectively.}
\label{M87_vlba}
\end{figure}

\begin{figure}
\centering{\includegraphics[width=10cm]{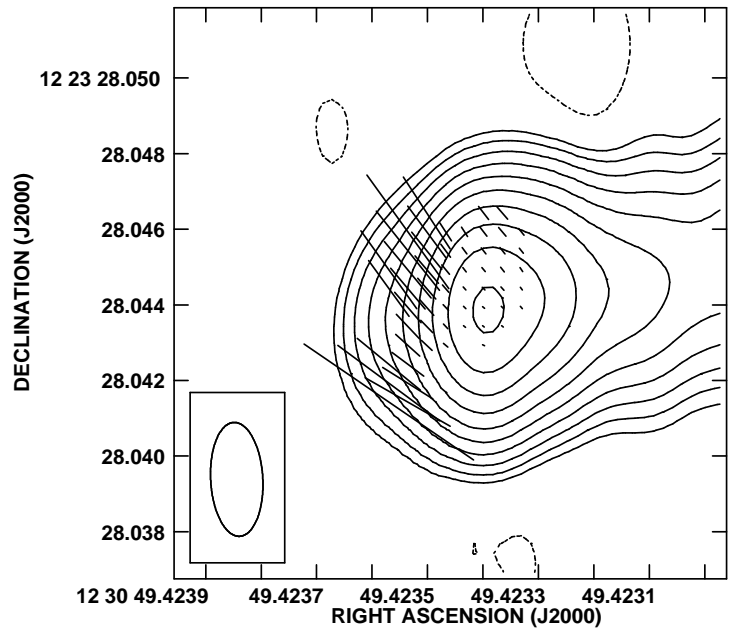}}
\caption{\small 5\,GHz VLBA image of the core in M\,87 with fractional polarization electric vectors superposed. 1 mas tick = 0.7\% polarization. The lowest contour is at
$\pm$0.17\% of the peak surface brightness of 1.06\,Jy\,beam$^{-1}$.} 
\label{M87_core}
\end{figure}

\begin{figure}
\centering{\includegraphics[width=8cm]{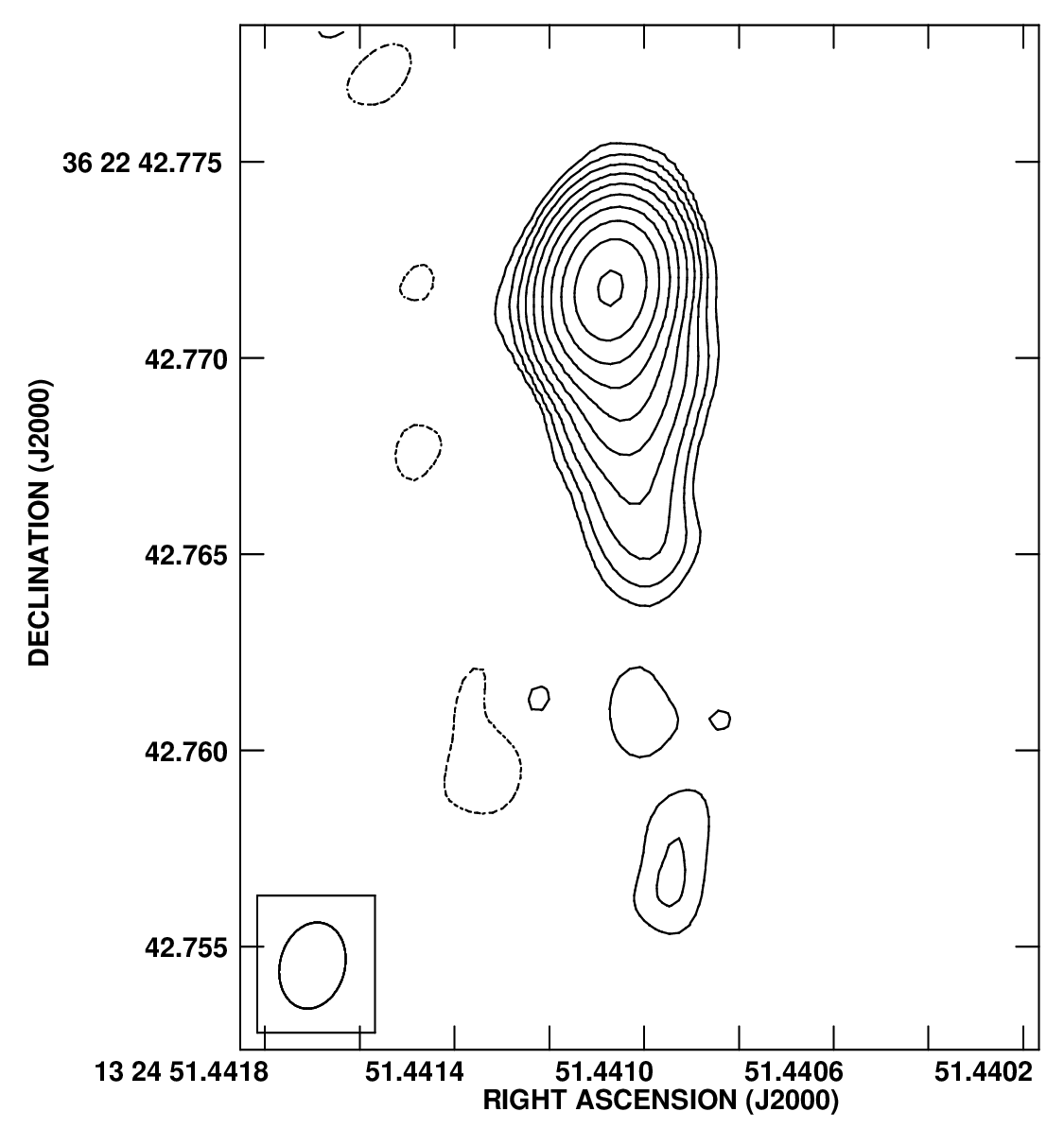}}
\caption[UGC\,08433: VLBA image]{\small 5\,GHz VLBA total intensity image of UGC\,08433. Polarization was not detected in this source. Contours are in percentage of peak surface brightness and increase in steps of 2: the lowest contour and peak surface brightness are $\pm$0.17\% and 55\,mJy\,beam$^{-1}$, respectively.}
\label{NGC5141_vlba}
\end{figure}

\begin{figure}
\centering{\includegraphics[width=8cm]{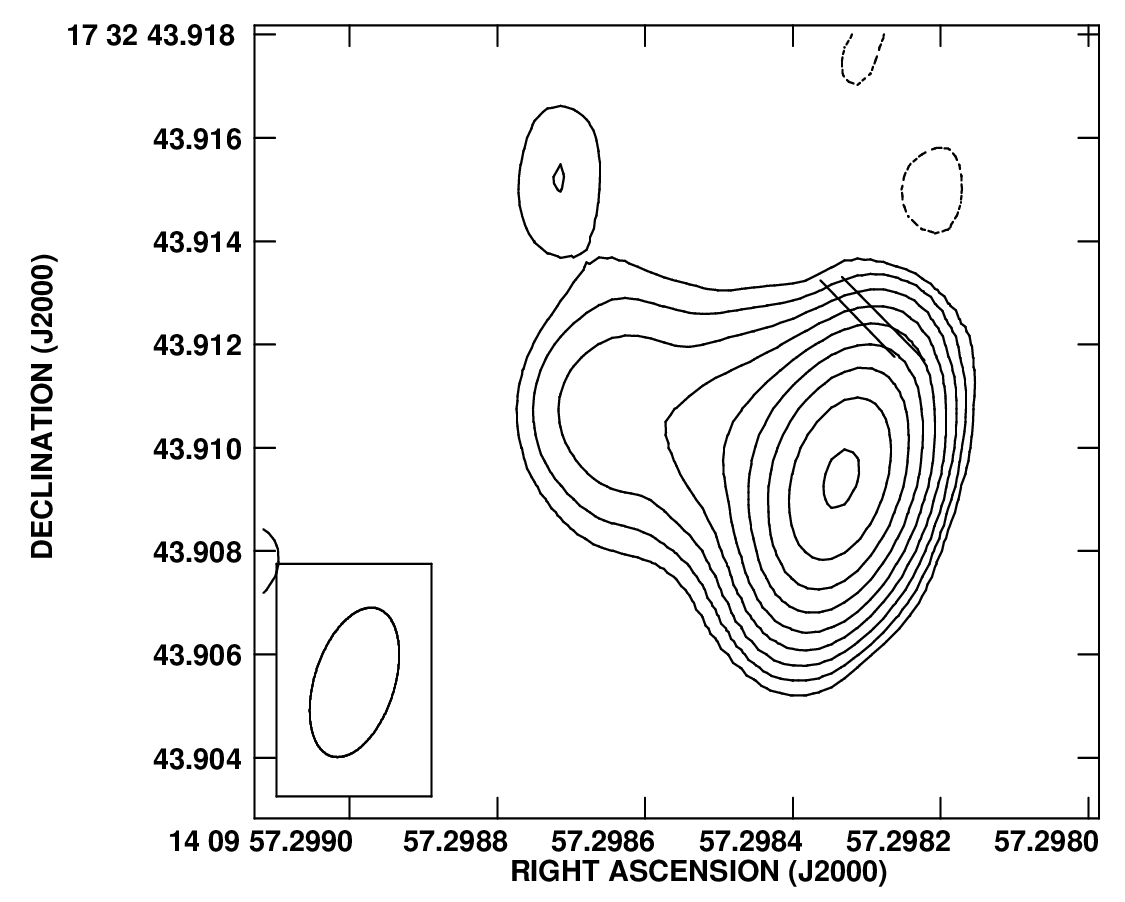}}
\caption[UGC\,09058: VLBA image]{\small 5\,GHz VLBA total intensity image of UGC\,09058 with fractional polarization electric vectors superposed. 1 mas tick = 13\% polarization. Contours are in percentage of peak surface brightness and increase in steps of 2: the lowest contour and peak surface brightness are $\pm$0.35\% and 22\,mJy\,beam$^{-1}$, respectively.}
\label{NGC5490_vlba}
\end{figure}

\begin{figure}
\centering{\includegraphics[width=8cm]{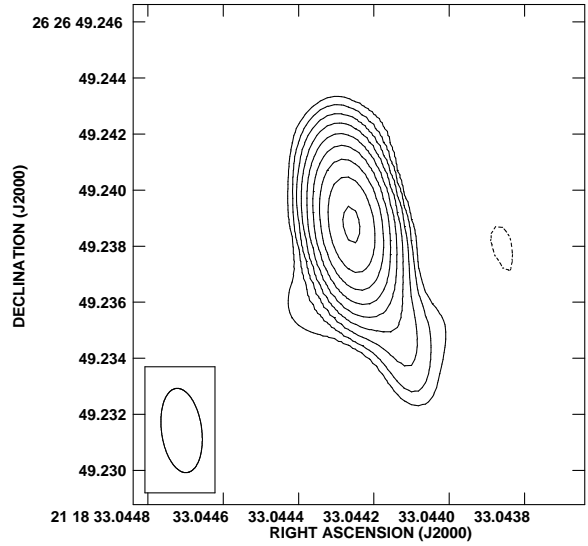}}
\caption[UGC\,11718: VLBA image]{\small 5\,GHz VLBA total intensity image of UGC\,11718. Polarization was not detected in this source. Contours are in percentage of peak surface brightness and increase in steps of 2: the lowest contour and peak surface brightness are $\pm$0.35\% and 36\,mJy\,beam$^{-1}$, respectively.}
\label{NGC7052_vlba}
\end{figure}

We present the 5\,GHz total intensity images with the fractional polarization vectors superimposed on them, for the 10 radio galaxies with ``core-jet'' structures in Figures~\ref{NGC193_vlba} to \ref{NGC7052_vlba}. The jets are captured in much greater detail {at 5\,GHz compared to 1.6\,GHz  \citep{xu00}: individual} jet components and bends are more prominently highlighted. The presence of one-sided ``core-jet'' structures is indicative of Doppler boosting due to relativistic {bulk motion in the jets} on parsec-scales. {We note that the Doppler boosting effect results in an overestimation of the ``equipartition'' magnetic fields, energies, and pressures, and an underestimation of the particle lifetimes listed in Table~\ref{equip}.} Of the 10 sources, all but three show signs of polarized emission in their parsec-scale jets. M\,87 was observed as the phase reference calibrator for M\,84. No core polarization is detected in any of the FRI radio galaxies, except in M\,87. Table~\ref{vlba_pol} summarizes the polarization properties. 

The total intensity estimates and other properties like the jet-to-counterjet surface brightness ratio, $R_{J}$, are presented in Table~\ref{vlba_tot}. We note that the estimates for $R_{J}$ are typically larger than those presented in \citet{xu00}. This is due to the different procedures adopted to derive them $-$ while \citet{xu00} divided the image with a 180$\degr$ rotated version of the same, and obtained average jet sidedness values from the resulting image, we have simply estimated the surface brightness along the center of the jet at a given distance from the core, and obtained the noise in the counterjet direction at approximately the same distance from the core. 

Three sources {\it viz.,} UGC\,06635, UGC\,07115, and UGC\,12064, were not detected in the 1.6\,GHz VLBA observations. These three were not observed in the first epoch observations of \citet{xu00}. The remaining six appear to be ``core-only'' sources. We did not detect any jet emission in the previously  ``core-jet'' classified source, UGC\,12531. Note that ``jet'' emission was merely hinted at in the previous observation by \citet{xu00}. Xu et al. had tentatively classified UGC\,11718 (NGC\,7052) as a ``twin-jet'' source at 1.6\,GHz, but we see a clear ``core-jet'' structure at 5\,GHz. It is possible that the counterjet in this source was an artifact due to errors in amplitude calibration, as was also recognized by Xu et al. 

\begin{figure}
\centering{\includegraphics[width=9cm]{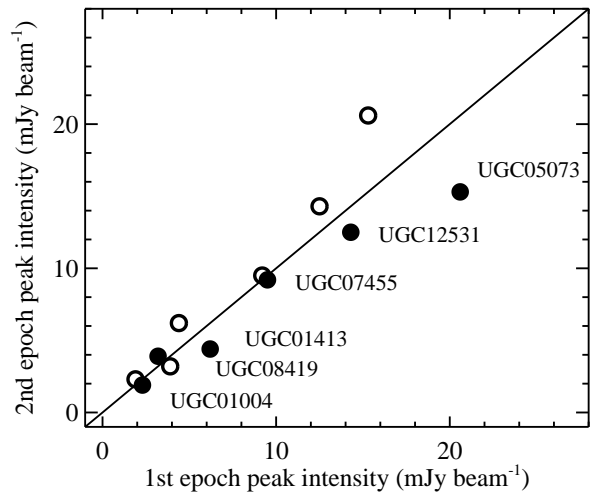}}
\caption{\small A comparison of the first epoch (filled circles) and second epoch (empty circles) 1.6\,GHz VLBA observations of six ``core-only'' UGC galaxies. The intensity errors are smaller than the symbol sizes.}
\label{variability}
\end{figure}

A comparison {of} the 1.6\,GHz core peak flux densities from the two epochs of observation for the six sources is made in Figure~\ref{variability}. The core flux appears to have varied on average by a factor of 1.2 over a period of 5.8 years. Only the cores in UGC\,01413 and UGC\,05073 exhibit a slightly larger {than average} variation, {\it i.e.,} by a factor of 1.4. This variation is consistent with the $\lesssim$2 factor variability typically observed in FRI radio galaxy cores \citep[e.g.,][]{Sadler94,Evans05}. The radio core prominence parameter, {$R_{c}$, which is a {\it statistical} indicator of beaming and thereby orientation,} is = 0.02 and 0.30 in these two sources, respectively \citep{Kharb04A}. While UGC\,05073 has a core prominence value that lies at the lower end of values exhibited by BL Lac objects \citep[see Figure 3 in][]{Kharb04A}, UGC\,01413 has a prominence value similar to that observed in a majority of FRI radio galaxies. {While keeping in mind the fact that $R_{c}$ 
must be cautiously used as an orientation proxy for individual sources,} we do not observe a clear trend of increased core variability and orientation in the small number of galaxies considered here. Higher dynamic range images are likely required to detect jets in these sources.



\begin{deluxetable}{cccccccccc}
\tabletypesize{\small}
\tablecaption{VLBA Observations at 5\,GHz: Total Intensity}
\tablewidth{0pt}
\tablehead{
\colhead{Source} &  \colhead{Phase}  &\colhead{$I_{peak}^{c}$} &\colhead{$I_{int}^{c}$}&\colhead{$S_{tot}$}&\colhead{$I_{rms}$} & \colhead{Beam} & \colhead{R$_{J}$} & \colhead{$d$} \\
\colhead{Name}&\colhead{Calibrator}&\colhead{mJy/beam}&\colhead{mJy}&\colhead{mJy}&\colhead{mJy/beam}&\colhead{mas$\times$mas}&\colhead{}&\colhead{mas}}
\startdata
UGC\,00408& J0040+0125 &34.5  &39.8   &43.7& 4.7E$-$2&  3.4$\times$1.6  & $>$26 & 3.0 \\
UGC\,00689& J0112+3208 & 71.0  &80.4   &85.9& 5.1E$-$2&  2.6$\times$1.7  & $>$23 & 1.3 \\
UGC\,01841& J0222+4302&137.2&165.4 &194.3& 4.5E$-$2&  2.3$\times$1.5 & $>$132 & 3.2\\
UGC\,03695& J0710+4732 & 49.3  &56.4    &61.6& 5.3E$-$2&  2.4$\times$1.4   & $>$29 & 1.5\\
UGC\,06723& J1148+1840 & 110.4&130.8 &159.3& 4.9E$-$2&  2.6$\times$1.7 & $>$198 & 1.9\\
UGC\,07494& J1230+1223 & 172.3&197.3 &208.1& 5.4E$-$2&  3.0$\times$1.4  & $>$28  & 0.4\\
UGC\,07654& ....$^{a}$ & 1040.4&1803.0 &2492.5& 3.6E$-$1&  3.0$\times$1.4  & $>$258 & 4.1 \\
UGC\,08433& J1317+3425 & 55.1  &62.0   &66.6& 4.8E$-$2&  2.2$\times$1.6   & $>$23 & 0.8\\
UGC\,09058& J1406+1859 & 21.8  &24.5   &26.3& 4.2E$-$2&  3.0$\times$1.5   & $>$33 & 2.4\\
UGC\,11718& J2114+2832& 36.1  &39.6   &40.7& 5.2E$-$2&  3.3$\times$1.4   & $>$20 & 1.9
\enddata
\tablecomments{Col.\,1: Source name. Col.\,2: Phase reference calibrator. ${a}~-$ UGC\,07654 was observed as the phase calibrator for UGC\,07494. Col.\,3 \& 4: Core peak and integral intensity, respectively, as measured with JMFIT. Col.\,5: Total source flux density as measured with TVWIN+IMSTAT. Col.\,6: Final {\it rms} noise in image. Col.\,7: Synthesized beam-size. Col.\,8 \& 9: Jet to counterjet surface brightness ratio, and the distance along the jet where this ratio was estimated. The {\it rms} noise in the image was used for the counterjet intensity.}
\label{vlba_tot}
\end{deluxetable}

\begin{deluxetable}{cccccccccc}
\tabletypesize{\small}
\tablecaption{2nd Epoch VLBA Observations at 1.6\,GHz: Total Intensity}
\tablewidth{0pt}
\tablehead{
\colhead{Source}&\colhead{Phase}&\colhead{$I_{peak}^{c}$} &\colhead{$I_{int}^{c}$}&\colhead{$I_{rms}$}\\
\colhead{Name}&\colhead{Calibrator}&\colhead{mJy/beam}&\colhead{mJy}&\colhead{mJy/beam}}
\startdata
UGC\,01004& J0125$-$0005& 2.3 $\pm$ 0.17  & 2.7 $\pm$ 0.33 & 0.22 \\ 
UGC\,01413& J0155+0438    & 6.2 $\pm$ 0.08  & 7.5 $\pm$ 0.17 & 0.09 \\
UGC\,05073& J0907+6815    & 20.6 $\pm$ 0.08 & 21.9 $\pm$ 0.14 & 0.07 \\
UGC\,06635& J1148+1840    & ... & ... & 0.05 \\
UGC\,07115& J1209+2547    & ... & ... & 0.12 \\
UGC\,07455& J1217+5835    & 9.5 $\pm$ 0.09 & 9.6 $\pm$ 0.16 & 0.08 \\
UGC\,08419& J1329+3154    & 3.2 $\pm$ 0.07 & 3.8 $\pm$ 0.13 & 0.07\\
UGC\,12064& J2241+4120    & ... & ... & 0.07 \\
UGC\,12531& J2322+0812    & 14.3 $\pm$ 0.13 & 14.7 $\pm$ 0.23 & 0.12\\ 
\enddata
\tablecomments{Col.\,1: Source name. Col.\,2: Phase reference calibrator. Col.\,3 \& 4: Core peak and integral intensity, respectively, as measured with JMFIT. Col.\,6: Final {\it rms} noise in image. {The images were convolved with circular beams of size 10~mas.}}
\label{vlba_sectot}
\end{deluxetable}

\begin{deluxetable}{cccclccl}
\tabletypesize{\small}
\tablecaption{Equipartition Estimates}
\tablewidth{0pt}
\tablehead{\colhead{Source} & \colhead{Length} & \colhead{Width}&  \colhead{L$_{rad}$} & \colhead{B$_{min}$} & \colhead{E$_{min}$} & \colhead{P$_{min}$} & \colhead{t$_{syn}$} \\
\colhead{Name} & \colhead{mas} &\colhead{mas} &\colhead{erg\,s$^{-1}$} &\colhead{milliG} &\colhead{ergs} &\colhead{dyn\,cm$^{-2}$} &\colhead{yrs}}
\startdata
UGC\,11718  & 5.3 & 3.5 & 3.57E+38 & 3.2 & 8.00E+49 & 9.33E$-$07 &  2045 \\
UGC\,03695  & 6.8 & 3.5 & 9.88E+38 & 3.1 & 2.32E+50 & 8.76E$-$07 &  2142 \\
UGC\,00689  & 4.3 & 3.5 & 1.25E+39 & 4.4 & 1.69E+50 & 1.82E$-$06 &  1238 \\
UGC\,00408  & 5.3 & 4.6 & 2.58E+38 & 2.6 & 7.73E+49 & 6.34E$-$07 &  2737 \\
UGC\,06723  & 4.3 & 4.6 & 8.79E+39 & 4.9 & 1.02E+51 & 2.24E$-$06 &  1059 \\
UGC\,08433  & 4.3 & 3.5 & 4.90E+38 & 3.0 & 1.15E+50 & 8.68E$-$07 &  2157 \\
UGC\,07654  & 7.5 & 6.3 & 2.20E+39 & 7.9 & 1.24E+50 & 5.84E$-$06 &  520   \\
UGC\,07494  & 4.3 & 4.6 & 9.38E+37 & 5.1 & 1.02E+49 & 2.45E$-$06 &  998   \\
UGC\,09058  & 4.3 & 3.5 & 8.81E+38 & 3.8 & 1.49E+50 & 1.35E$-$06 &  1549 \\
UGC\,01841  & 6.8 & 5.6 & 7.27E+39 & 4.0 & 1.14E+51 & 1.50E$-$06 &  1430 \\
\enddata
\tablecomments{Col.\,1: Source name. Col.\,2 \& 3: Length and (deconvolved) jet width used for the estimation $-$ a small portion of the jet a few mas from the core {with roughly constant brightness} was chosen to get the estimates. Col.\,4: Total radio luminosity. Col.\,5: ``minimum'' magnetic field strength. Col.\,6 \& 7: ``minimum'' energy and pressure. Col.\,8: Synchrotron lifetime of electrons undergoing both radiative and inverse Compton (on CMB photons) losses, for a break frequency of 5 GHz. }
\label{equip}
\end{deluxetable}

\begin{deluxetable}{ccccccccc}
\tabletypesize{\small}
\tablecaption{VLBA Observations at 5\,GHz: Polarized Intensity}
\tablewidth{0pt}
\tablehead{
\colhead{Source} & \colhead{$P_{peak}$} &  \colhead{$P_{rms}$} &\colhead{$d$} & \colhead{$\chi_{mean}$} & \colhead{$\Delta{\chi}$} & \colhead{$m_{mean}$} & \colhead{$\Delta{m}$}\\
\colhead{Name}&\colhead{mJy/beam}&\colhead{mJy/beam}&\colhead{mas}&\colhead{degree}&\colhead{degree}&\colhead{$\%$}&\colhead{$\%$}}
\startdata
UGC\,00408 & ...      & 4.14E$-$2 &...& ... & ... & $<$5 & ...\\
UGC\,00689 & 0.25 & 4.75E$-$2 &10& 52&9& 56&17 \\
UGC\,01841 & 0.21 & 4.23E$-$2 &10& $-$47& 9 & 25 & 8\\
UGC\,03695 & 0.24 & 5.01E$-$2 &4& 81 & 9 & 13 & 4\\
UGC\,06723 & 0.79 & 4.48E$-$2 &4& 9 & 8 & 5 & 1 \\
                        &          &                &6& 63 & 5 & 13 & 2 \\
                        &          &                &8& 54 & 6 & 13 & 3 \\
                        &          &                &14& 6 & 9 & 22 &  7\\
UGC\,07494 & 0.31 & 5.03E$-$2 &3& $-$83 & 8 & 25 & 7\\
UGC\,07654 & 0.88 & 8.39E$-$2 &core& 42 & 6 & 0.13 & 0.02\\
                        &          &                &$<3$, cjet& 43 & 7 & 0.9 & 0.2\\
                        &          &                &7& 0 & 8   &  5 & 1\\                          
                        &          &                &9& 45 & 7 & 5 & 1\\
                        &          &                &36& $-$51 & 9 & 19 & 8\\                        
UGC\,08433 & ...      & 4.46E$-$2 &...& ... & ... & $<$4 & ...\\
UGC\,09058 & 0.32 & 7.66E$-$2 &3& 46 & 8 & 28 & 7\\
UGC\,11718 & ...      & 4.29E$-$2 &...& ... & ... & $<$4 & ...\\
\enddata
\tablecomments{Col.\,1: Source name. Col.\,2: Peak polarized intensity. Col.\,3: Final {\it rms} noise in polarization map. 
{Col.\,4: Distance of polarized jet component from the core. Polarization was detected from several jet components in UGC\,06723 and UGC\,07654. ``cjet'' for UGC\,07654 or M\,87 refers to polarized emission from the ``counterjet'' region.}
Col.\,5 \& 6: Mean polarization angle and error, respectively. Col.\,7 \& 8: Mean fractional polarization and error, respectively. Limits for jet fractional polarization for sources without a detection are listed in Col.\,7.}
\label{vlba_pol}
\end{deluxetable}

\subsection{Notes on Individual Sources with Polarized Emission}
{\bf UGC\,00689}: 
Polarization is detected mainly along the jet edge at a distance of $\sim10$\,mas in UGC\,00689. The polarization electric vectors are transverse to the jet direction. As the inferred magnetic field for optically thin emission is transverse to the polarization electric vectors, this implies that the magnetic field is largely aligned ($B_{||}$) at the jet edge. This in principle could support the  scenario of jet-medium interaction which ``stretches'' the magnetic field lines due to ``shear'' at the jet edges, giving rise to an aligned magnetic field \citep[e.g.,][]{Attridge99}. However, such field geometries can also rise from helical magnetic fields or magnetic fields with a dominant toroidal component threading these jets \citep[e.g.,][]{Kharb09,Clausen11}. It is interesting to note that the jet in this source has the highest fractional polarization ($56\%\pm17\%$) in the sample, approaching the theoretical maximum for optically thin incoherent synchrotron radiation. This is consistent with the magnetic field being highly ordered on parsec scales. Such a high degree of polarization ($\approx60\%$) has been observed in some blazar jets \citep[e.g.,][]{Roberts90,Cawthorne93}.

{\bf UGC\,01841}: Polarization is detected largely along the jet edge at a distance of $\sim10$\,mas in UGC\,01841. The polarization electric vectors are transverse to the jet direction implying an aligned magnetic field ($B_{||}$) at the jet edge. 

{\bf UGC\,03695 (NGC\,2329)}: 
This source shows marginal polarization in its ``inner'' jet ($\sim4$\,mas from the core). Assuming optically thin emission, the inferred magnetic field is oblique to the inner jet direction.

{\bf UGC\,06723 (3C\,264)}: The VLBA image shows an extended parsec-scale jet aligned with the kiloparsec scale VLA jet. The VLBA jet extends $\sim20$\,mas from the core. Significant polarization is detected in the parsec-scale jet at mostly $\sim$6\,mas from the core. The polarization vectors in this region change from nearly transverse to the jet direction at the jet edge to nearly aligned with the jet in the jet center. The inferred magnetic field is nearly parallel ($B_{||}$) at the jet edge, to nearly perpendicular ($B_{\perp}$) towards the jet center. There is also a region at $\sim4$\,mas from the core (``inner'' jet) where the polarization electric vectors are oblique to the jet direction. Such a structure is also observed at $\sim14$\,mas from the core. The complex magnetic field structure in the region closer to the core is reminiscent of the ``spine-sheath'' $B$-field structure that is sometimes observed in parsec-scale jets \citep[e.g.,][]{Attridge99}. The slight offset from a purely perpendicular or parallel magnetic field structure in the jet of UGC\,06723 could be due to Faraday rotation from the surrounding ionized gas. A rotation measure (RM) of $\sim50$~rad\,m$^{-2}$ would be required to rotate the electric vectors by 10$\degr$ at 5\,GHz.

{\bf UGC\,07494}: There is a hint of polarized emission close to the core of UGC\,07494 and is most likely originating in the ``inner'' jet region. The polarization seems to be in the counterjet region and perpendicular to the overall jet direction ({\it i.e.,} inferred $B_{||}$ field). 

{\bf UGC\,07654 (M\,87)}: This source shows a jet structure extending to nearly 40\,mas from the core. Polarization is detected in the core-counterjet region with the polarization electric vectors aligned obliquely to the jet direction. Some weak polarization (fractional polarization of a few percent) is detected at the jet edges around 10\,mas from the core, transverse to the jet direction ({\it i.e.,} inferred $B_{||}$ field). Polarization is also detected at a distance of $\sim$36\,mas from the core with an inferred $B_{\perp}$ field to the jet direction. {M\,87 has been the subject of several VLBI studies \citep[e.g.,][]{Junor99,Asada12}. Consistent with our detection of core polarization in M\,87, preliminary results from the 43\,GHz VLBA observations of Craig Walker and collaborators, indicate polarization in the core-counterjet region of this source.\footnote{See the online proceedings of the NRAO-NAASC 2012 Workshop titled ``Outflows, Winds and Jets: From Young Stars to Supermassive Black Holes''.}}

{\bf UGC\,09058}: {Polarized emission is detected in the ``inner'' jet region, $\sim3$\,mas from the core, in UGC\,09058.} The oblique polarization electric vectors could be suggestive of Faraday rotation due to {an} ionized medium around the core region. They could also suggest the presence of oblique shocks.

\section{Chandra Observations \& Data Analysis} 
Of the 21 sources in {the} sample, eight sources have published images with the AXAF CCD Imaging Spectrometer (ACIS) on board the {\it Chandra} space telescope  \citep{Evans06,Sambruna03,Croston07}. The X-ray images of UGC\,00597, UGC\,00689, UGC\,01841, UGC\,06723, UGC\,07360, UGC\,07494 and UGC\,07654 are also  presented on the XJET webpage\footnote{http://hea-www.harvard.edu/XJET/}. \citet{Croston07} describe the X-ray observations of UGC\,06635. In addition, archival data exists for six other sources, {\it viz.,} UGC\,01004, UGC\,01413, UGC\,03695, UGC\,08433, UGC\,11718 and UGC\,12064. We obtained new $\sim30$\,ks {\it Chandra}-ACIS imaging data in the VFAINT mode for UGC\,00408. Of the seven sources with archival or new data, UGC\,01413, UGC\,11718 and UGC\,12064 do not show any distinctive jet-like features and moreover do not have sufficient counts in their images for a robust spectral fit. Overall, nine of the 15 UGC galaxies studied with {\it Chandra} exhibit X-ray jets in them. We focus now on the {\it Chandra} observations of UGC\,00408 and UGC\,08433.

\begin{deluxetable}{cclccl}
\tabletypesize{\small}
\tablecaption{X-ray data}
\tablewidth{0pt}
\tablehead{
\colhead{Source}&  \colhead{OBSID}&
\colhead{Observer}& \colhead{Date}& \colhead{Exposure}& \colhead{Instrument}}
\startdata
UGC\,00408& 4053 & O'Dea& 2003-09-01& 29.44 ks& ACIS\\
UGC\,08433& 4055 & Sambruna& 2003-11-25& 30.97 ks& ACIS\\
\enddata
\label{archive_data}
\end{deluxetable}

Starting with level 1 event files, we removed the position randomization and applied the appropriate corrections for gain and charge transfer efficiency. Level 2 event files were obtained by following appropriate steps described in {the} {\it Chandra} data reduction threads. We used {the} ``Subpixel Event Repositioning'' (SER) technique described by \citet{li03,li04}, to improve {the} image resolution. Spectra were extracted from different regions for each of the sources. The data were binned so that each bin had at least 20 counts. The spectra of the core and jet structures were extracted and analyzed separately. The core spectrum was extracted from a circular region of radius 1$\asec$ centered on the brightest pixel. An aperture of this size includes $\approx$90\% of the total flux at 1~keV from a point source from that region. Jet spectra were extracted from rectangular regions adjusted to minimize non-jet emission. However, there were insufficient counts for an independent fit for the jet spectra for either source.

\subsection{Results from the X-ray study}
\begin{figure}
\centering{
\includegraphics[width=9.2cm,trim=2cm 3cm 0cm 4cm]{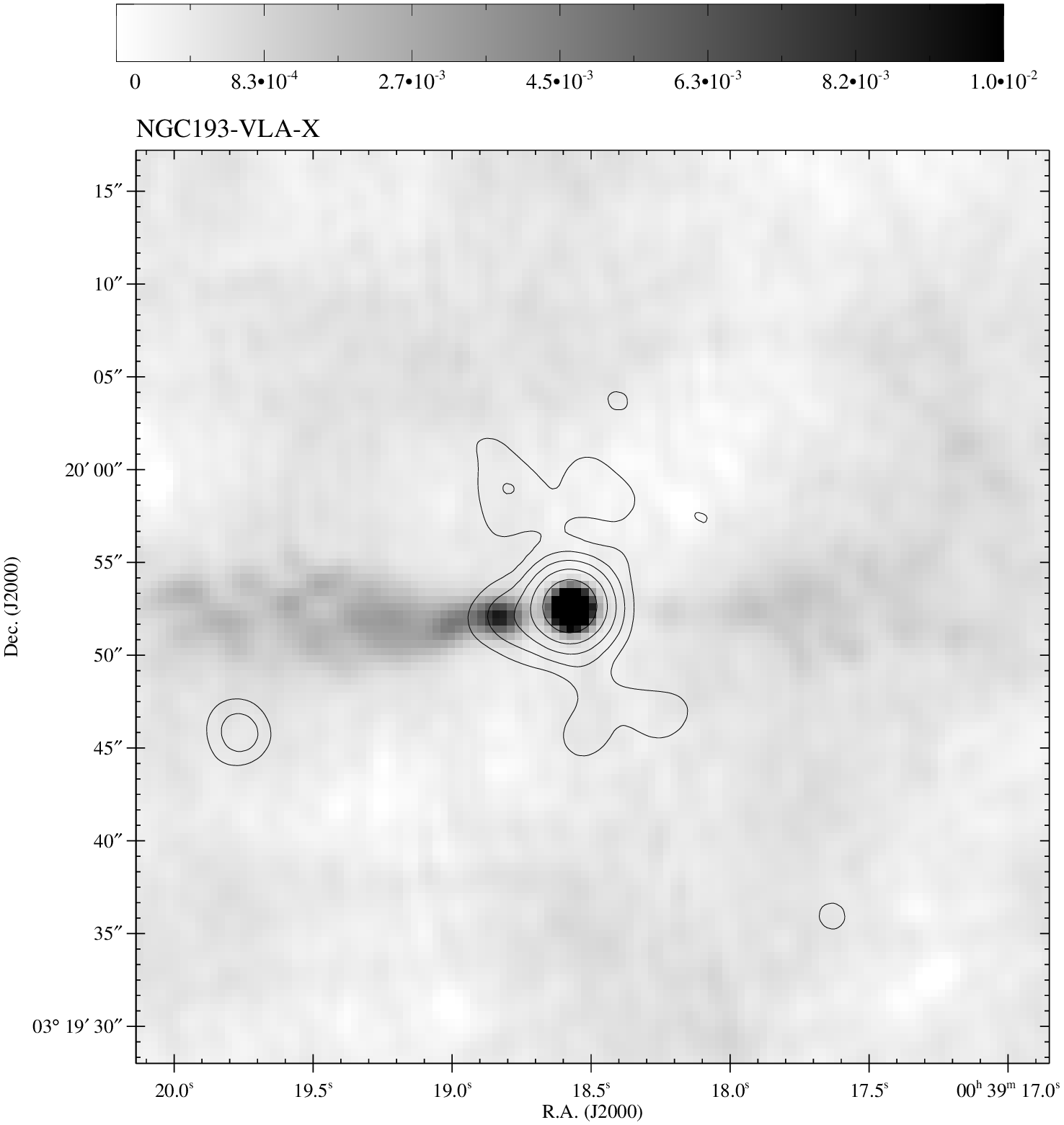}
\includegraphics[width=8cm,trim=2cm 0.5cm 0cm 0cm]{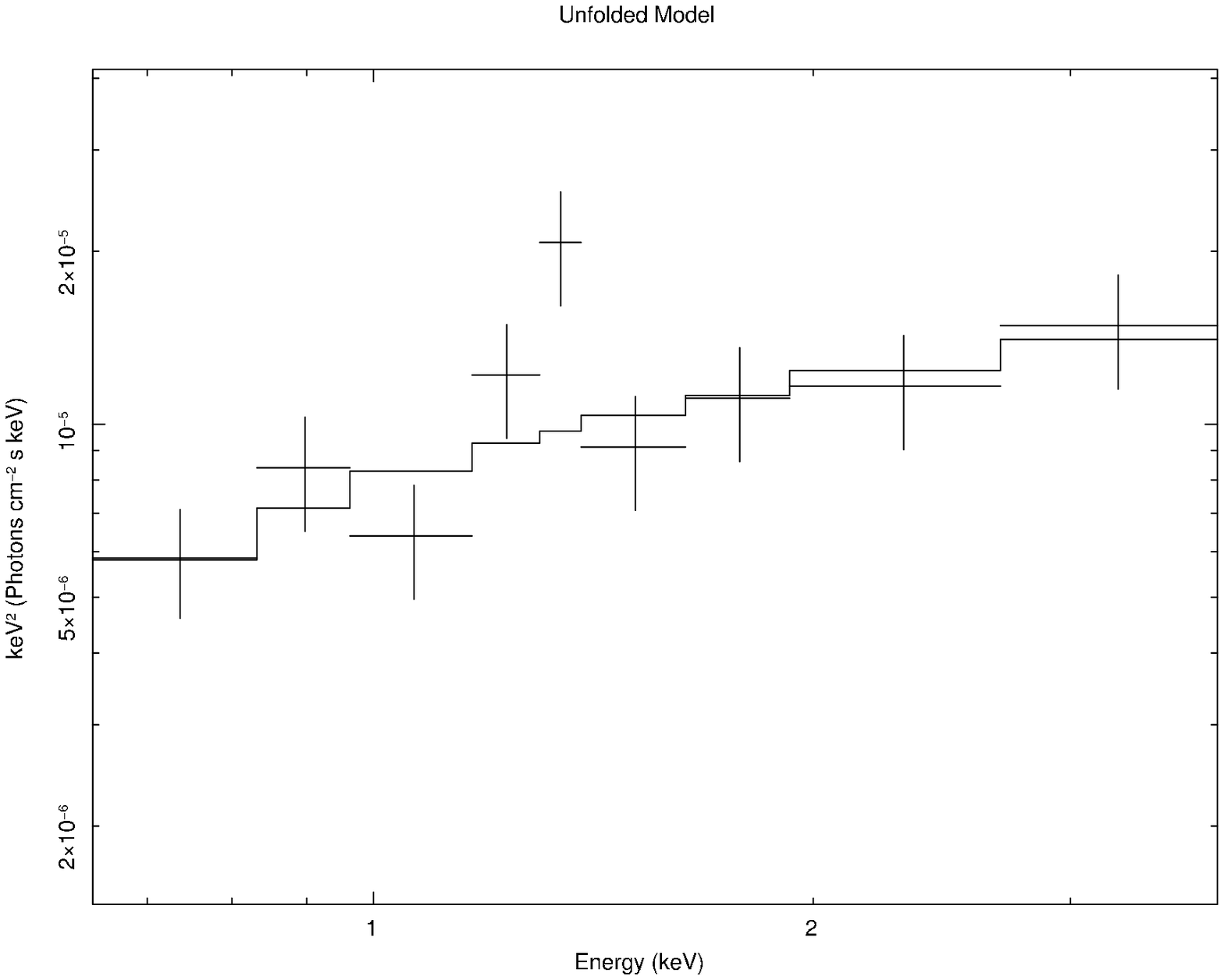}}
\caption{\small (Top) The overlay of UGC\,00408 {or NGC\,193} with {the} X-ray image in contours and {the} radio image in greyscale. The contour levels are at 3$\sigma$$\times$[1, 2, 4, 8, 16, 32, 64, 128, 256]. (Bottom) Spectral distribution of the X-ray emission in the nuclear region of UGC\,00408. The Figure also shows the unfolded model contributiong to the final spectrum.}\label{NGC193_overlays}
\end{figure}

\begin{figure}
\centering{\includegraphics[width=8cm]{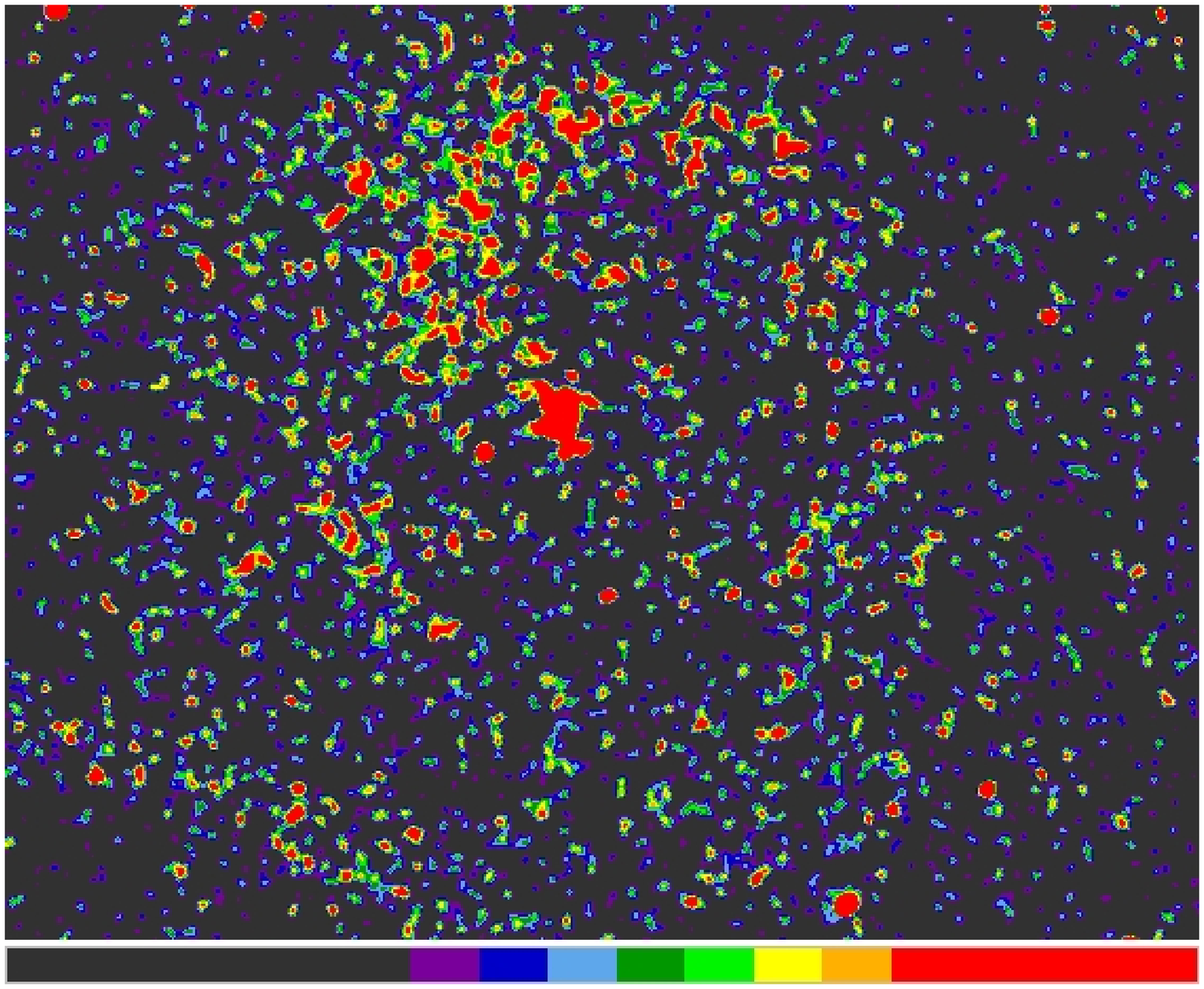}}
\caption{\small Color image highlights large scale diffuse emission detected approximately 1$\arcmin$ from the nucleus in UGC\,00408. The X-ray image has been smoothed with a Gaussian of width = 2 pixels.}
\label{NGC193_ring}
\end{figure}

{\bf UGC\,00408 (NGC\,193):}
This S0 galaxy is part of a galaxy group and seems to have a companion, NGC\,204. 
The {\it Chandra}-ACIS observations reveal X-ray emission along the radio jet (Figure \ref{NGC193_overlays}), extending approximately 5$\asec$ ($\sim$1.5 kpc). The nuclear spectrum can be adequately described by a simple powerlaw, with absorption consistent with the Galactic column. The number of counts was too low to justify more complex modeling. 
The distribution of the large scale diffuse emission can be seen more clearly in Figure \ref{NGC193_ring}. {By smoothing the {\it Chandra} image with a Gaussian of kernel radius 20 in DS9, we could clearly discern the presence of a bubble-like structure centered around the AGN, with a maximum outer radius of $\approx107\arcsec$ (=29 kpc). 
We extracted counts from this region, after excluding the inner region of about 19$\arcsec$ where the AGN emission dominates, using the CIAO software version 4.3 (CALDB version 4.4.6). We fitted a thermal gas model to the diffuse emission using the XSPEC software (model = wabs*MEKAL). The neutral hydrogen column density was frozen to the Galactic value in the direction of NGC\,193 (=2.79E+22 cm$^{-2}$). The modeling yielded a best-fit temperature of 0.80$\pm$0.01 keV for the X-ray gas, and a reduced $\chi^{2}$ value of 1.12 for 237 degrees of freedom. The unabsorbed 0.5$-$7 keV X-ray flux from the entire bubble-like structure is $\sim$4.4E$-$13~erg~cm$^{-2}$~s$^{-1}$.
We estimated the work (W) done on the gas by the inflation of the bubble using the simple relation, W = nkTV, where n was the hydrogen density, assumed to be 0.1~cm$^{-3}$ \citep[e.g.,][]{Mathews03}, and V was the volume of the entire bubble (not just the X-ray cavity; see below), assumed to be spherical and uniformly filled with the 0.8 keV X-ray gas. This work   turns out to be $\approx$3.9E+59~erg, which could be regarded as an upper limit to the total energy expended by the AGN in creating the bubble. Considering only the region of the X-ray cavity (i.e., the region with a paucity of X-rays around the central AGN), which has a smaller radius of about 60$\arcsec$ (=16.4 kpc), the PV work done by the AGN in inflating the cavity is $\approx$6.9E+58~erg. Deeper {\it Chandra} observations of NGC\,193 have since been obtained, and will be discussed by Jones et al 2012, in prep.}


\begin{figure}
\centering{
\includegraphics[width=9.2cm,trim=2cm 3cm 0cm 4cm]{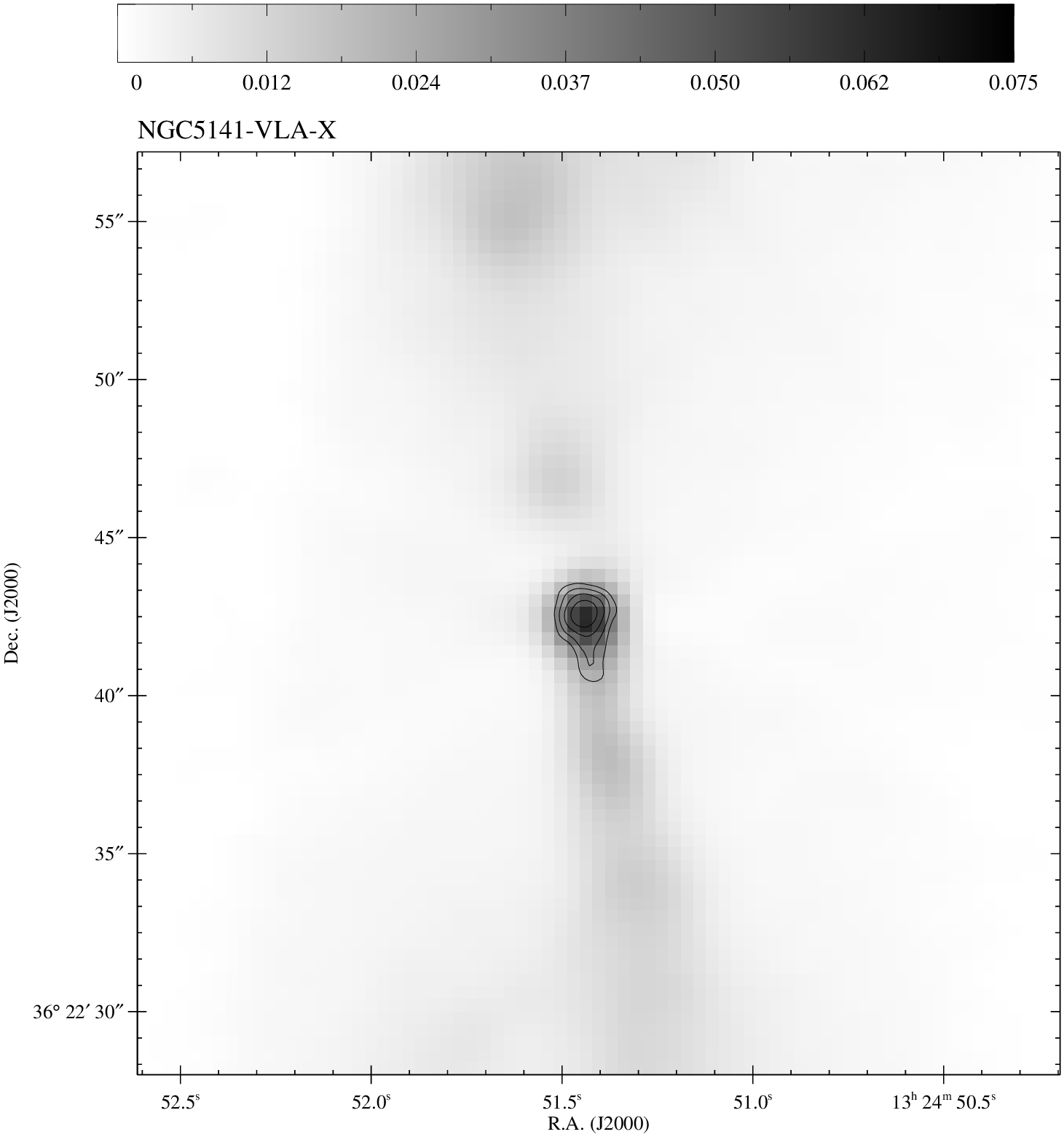}
\includegraphics[width=8cm,trim=2cm 0.5cm 0cm 0cm]{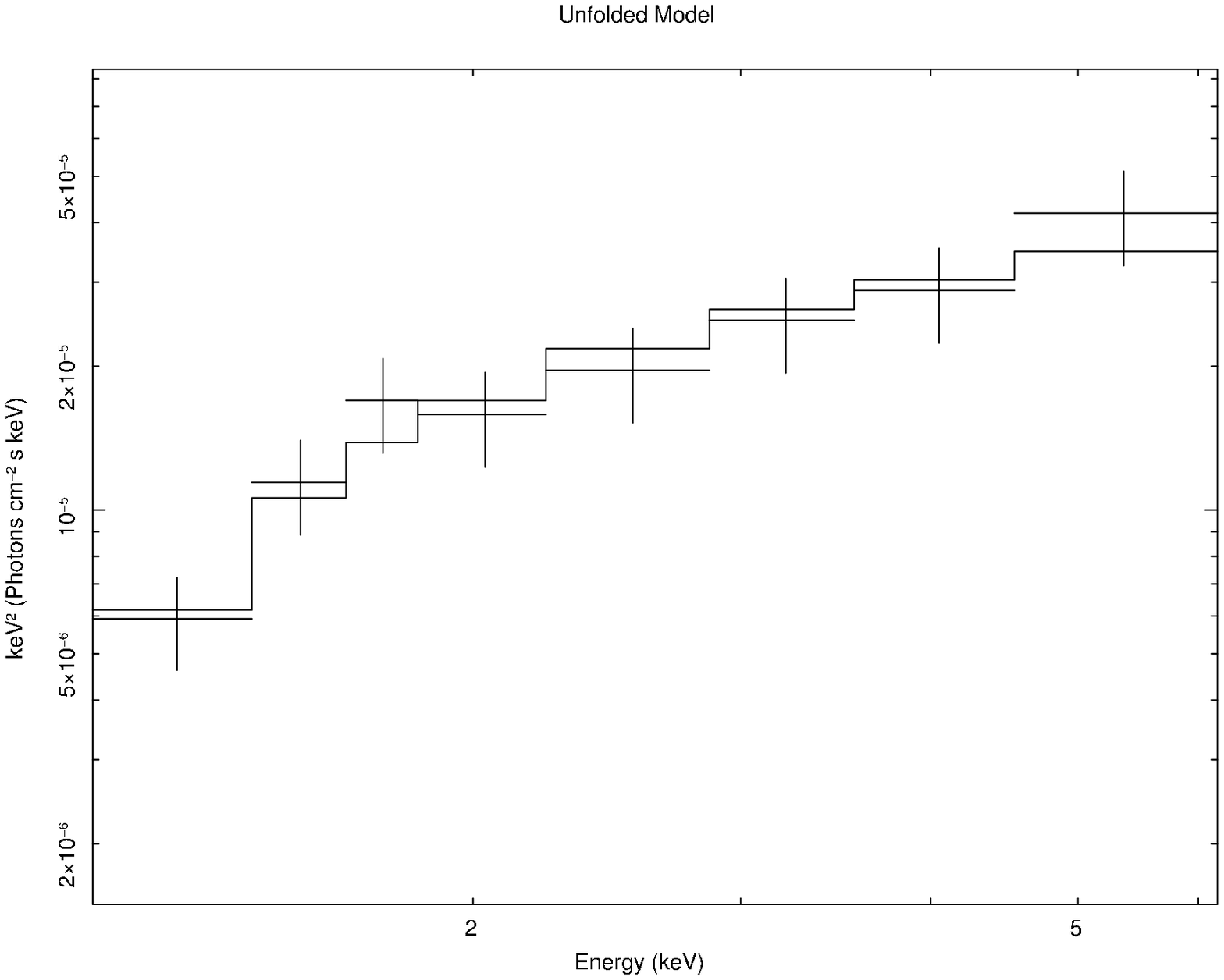}}
\caption{\small (Top) The overlay of UGC\,08433 {or NGC\,5141} with {the} X-ray image in contours and {the} radio image in greyscale. The contour levels are at 3$\sigma$$\times$[1, 2, 4, 8, 16, 32, 64, 128, 256]. (Bottom) Spectral distribution of the X-ray emission in the nuclear region of UGC\,08433. The Figure also shows the unfolded model contributing to the spectrum.}
\label{NGC5141_overlays} 
\end{figure}

{\bf UGC\,08433 (NGC\,5141):}
This S0 galaxy is a part of a group of six galaxies and has a known companion, NGC\,5142 \citep{Ramella89}. The {\it Chandra} image shows X-ray emission associated with the radio jet (Figure \ref{NGC5141_overlays}). The X-ray jet is approximately 4$\asec$ ($\sim$1.5 kpc) long. There also seems to be some diffuse emission around the nucleus. The details of spectral parameters obtained during fitting are described in Table~\ref{Xray_spec}. The nuclear spectrum is adequately described by a simple powerlaw. The absorption column density is significantly higher than the Galactic value in that direction, and absorption due to the host galaxy may be significant. However, the nucleus does not have a sufficient S/N to fully characterize the spectrum. 

\section{Discussion}
It was concluded by \citet{xu00} that adiabatic expansion losses in a jet with a constant jet velocity and opening angle and a primarily transverse magnetic field could explain the surface brightness ``dimming" in the VLBA jets of the UGC radio galaxies. In order to resolve the jets and examine the magnetic field structure, we undertook higher resolution polarization-sensitive observations, which are presented here. The ``core-only" sources were observed for a second epoch at 1.6\,GHz to look for possible jet emission. In addition, two sources that had been missed being observed earlier, were looked at. However, these second epoch {1.6\,GHz} observations failed to detect jet emission, and the two new targets failed to get detected by the VLBA. The polarimetric observations did reveal polarized emission in seven of the ten sources observed, in jet knots a few milliarcsec away from the core.

\begin{deluxetable}{cclcc}
\tabletypesize{\small}
\tablecaption{X-ray Spectral analysis: UGC\,00408 and UGC\,08433}
\tablewidth{0pt}
\tablehead{
\colhead{Source}& \colhead{XSPEC} &\colhead{Best-fit} & \colhead{Flux 1\,keV} & \colhead{Luminosity 1\,keV}\\
\colhead{Name}&\colhead{Model}&\colhead{Parameters}&\colhead{erg cm\mtwo s\mone keV\mone}&\colhead{erg s\mone keV\mone}}
\startdata
UGC\,00408&  phabs*powerlaw& $N_{H}$ = $8.46^{+0.11}_{-0.08}$$\times$10$^{20}$ cm\mtwo& 1.8$\times$10$^{-14}$& 7.8$\times$10$^{39}$\\
& & $\alpha$ = $0.68_{-0.52}^{+0.55}$&&\\
& & $\chi^2$/d.o.f. = 9.43/6 &&\\
UGC\,08433&  phabs*powerlaw& $N_{H}$ = $8.69^{+6.29}_{-8.28}$$\times$10$^{21}$ cm\mtwo&2.9$\times$10$^{-14}$& 1.3$\times$10$^{40}$\\
& & $\alpha$ = $0.62_{-0.60}^{+0.54}$&&\\
& & $\chi^2$/d.o.f. = 9.43/6&& 
\enddata
\tablecomments{Col.\,3: $N_{H}$ = the hydrogen column density. Col.\,4 \& 5: Aperture corrected flux density and luminosity density, respectively, at 1\,keV.}
\label{Xray_spec}
\end{deluxetable}

\subsection{Polarization Detection: Depolarizing Medium}
Polarization traces the orderliness and orientation of the magnetic field in a source. If the magnetic field is tangled, the polarization may change on scales too small to be resolved by the telescope, giving rise to ``beam'' depolarization \citep[e.g.,][]{lai80}. Another factor affecting the observed degree of polarization could be large amounts of ionized gas in the central region. Extended X-ray emission is detected in a large number of FRI sources observed with the {\it Chandra} X-ray Observatory \citep[e.g.,][]{Mathews03}. Ionized gas {containing a magnetic field} would cause Faraday rotation of the polarized light: if the regions with different electric vector position angles are not resolved by the telescope, ``beam" depolarization would occur. If the Faraday rotating medium is mixed with the synchrotron emitting electrons, then Faraday rotation occurring at various depths of the source would result in ``internal" depolarization \citep{Burn66,Cioffi80}. We briefly explore candidates for the Faraday rotating medium for the core and jet components below. 

\subsubsection{Core Polarization}
Polarization is not detected in the cores of 9 of the 10 FRI sources. {Only the core region of M\,87 exhibits polarized emission, consistent with the preliminary results of Craig Walker et al. (2012, in prep.).} The lack of polarization in almost all the sources could be signifying the presence of ionized gas on parsec-scales in these sources. The inner ionized edge of the tori could be a suitable candidate for {the Faraday rotating} medium. The broad line clouds, or the intercloud medium \citep[e.g.,][]{ODea89} could be other candidates. However, if it is supposed that FRIs lack obscuring tori and a significant broad line region, as proposed on the basis of the high detection rate of optical nuclei in HST images \citep[e.g.,][]{chia99,Kharb04A}, then this lack of polarization could either suggest disordered magnetic fields (however see the discussion on sensitivity below), the presence of diffuse ionized gas in the nuclear regions and/or beam depolarization. {The fact that core polarization is detected only in the nearest, most well resolved source, M\,87, is consistent with the latter suggestion.}

For the sample source UGC\,07360 (3C\,270), \citet{Kharb05} detected a core
fractional polarization of 0.4\%. Since Kharb et al. had used a global VLBI array including the 100m Effelsberg antenna, it is possible that the lower sensitivity of the current data plays a role in the non-detections. We note that the typical core fractional polarization observed in radio galaxies is $\le1\%$ \citep{Rudnick86,Kharb03,Kharb05,Kharb04}.
\citet{glio03} found that the {\it Chandra} and {\it XMM} data of 3C\,270 were best fit by a Compton thick absorber ($N_{H}\sim5\times10^{22}$ cm$^{-2}$) with a covering factor of $\sim$80\%. Therefore, beam depolarization and inadequate sensitivity could be important factors in the lack of core polarization in these galaxies.

\subsubsection{Jet Polarization}
We detect polarization in some parts of the parsec-scale jets of 7 out of 10 FRI radio galaxies. 
The fractional polarization in the jets is typically between 5\% and 25\% (Table~3). {The jet segment in UGC\,00689 has the highest fractional polarization ($56\%\pm17\%$) in the sample, close to the theoretical maximum for optically thin incoherent synchrotron radiation, and consistent with a highly ordered magnetic field on parsec scales.} The inferred magnetic field is mostly aligned with the jet direction in four of the seven sources, {\it viz.,} UGC\,00689, UGC\,01841, UGC\,07499, UGC\,07654 (M\,87). It appears to be oblique to the jet direction along the spine of UGC\,06723 (3C\,264), but more aligned at the edges. Such a ``spine-sheath'' structure can either be explained as shocks along the jet spine, and shear along the edges due to jet-medium interaction, or simply by {the presence of} a helical magnetic field \citep{Lyutikov05,Kharb08,Kharb08a}. 

Oblique polarization electric vectors are also observed in UGC\,03695 and UGC\,09058. These could arise due to Faraday rotation by an ionized screen. At 5\,GHz, an RM of $\sim$218 rad\,m$^{-2}$ would be required to rotate the EVPA by 45$\degr$. We note that when RM is actually measured in FRI radio galaxy jets through multi-frequency polarimetric observations, it turns out to be the order of a few 100 rad\,m$^{-2}$ \citep[e.g.,][]{tay01,zav02,Kharb09}. \citet{Kharb09} reasoned that such an RM could result from a thin sheath (with a path length of say, $\sim$10\% of the jet width) surrounding the main synchrotron emitting body of the jet. While narrow line clouds on parsec-scales could also give rise to these RM, the short time-scale RM variability (over a few months) observed in a few sources, questions the suitability of NLR as a candidate \citep{Asada08}. Hot X-ray emitting gas observed with {\it Chandra} is a possible candidate for the Faraday rotating medium. However, the large path lengths involved for the polarized light, could result in large values for RM (several 1000 rad\,m$^{-2}$), which would then necessitate many reversals in the magnetic field direction to give final RM values of a few 100 rad\,m$^{-2}$. This would result in net depolarization at lower radio frequencies, an effect that is not observed in radio galaxy studies with multifrequency observations \citep[e.g.,][]{Kharb09}. But since this effect cannot yet be confirmed or {rejected} in {the} sample sources, the hot diffuse gas cannot be ruled out as a candidate for Faraday rotation. 

We detect significant fractional polarization in the jet of UGC\,06723 or 3C\,264. The degree of polarization is about $22$\%$\pm7$\% in the jet knot $\sim$15 mas from the core, and has an average value of $13$\%$\pm2.5$\% over the entire length of the jet. However, the jet also appears one-sided on kiloparsec-scales \citep{bau97} suggesting a small inclination angle. Thus the high polarization observed may be due to favorable orientation which Doppler boosts the radio emission. This can happen if individual jet components move with different Lorentz factors, so that the component with the faster speed dominates the polarized emission at smaller angles to line of sight \citep[e.g.,][]{Narayan11}. 

As we discuss below, the 5\,GHz observations are unable to resolve the jet structure in the transverse direction. Therefore, we conclude that beam depolarization is the major cause of the lack of polarization in the parsec-scale jets of these FRI radio galaxies. This is consistent with the finding that in the observations of 3C\,264 by \citet{Kharb09}, the average fractional polarization for a jet knot $\approx$2 mas from the core, which displayed an EVPA of $\sim$45$\degr$ {with respect to} the jet direction (see also Figure~\ref{3C264_vlba}), increased from $2.5$\%$\pm0.5$\% at 5~GHz to $5$\%$\pm0.7$\% at 8~GHz. As RM gradients resulting from helical magnetic fields, or Faraday rotation from hot ionized gas, could both be occurring in these parsec-scale jets, adequate resolution (e.g., at 15 GHz) {will be} crucial for the detection of jet polarization.

\begin{figure}
\includegraphics[width=8.5cm]{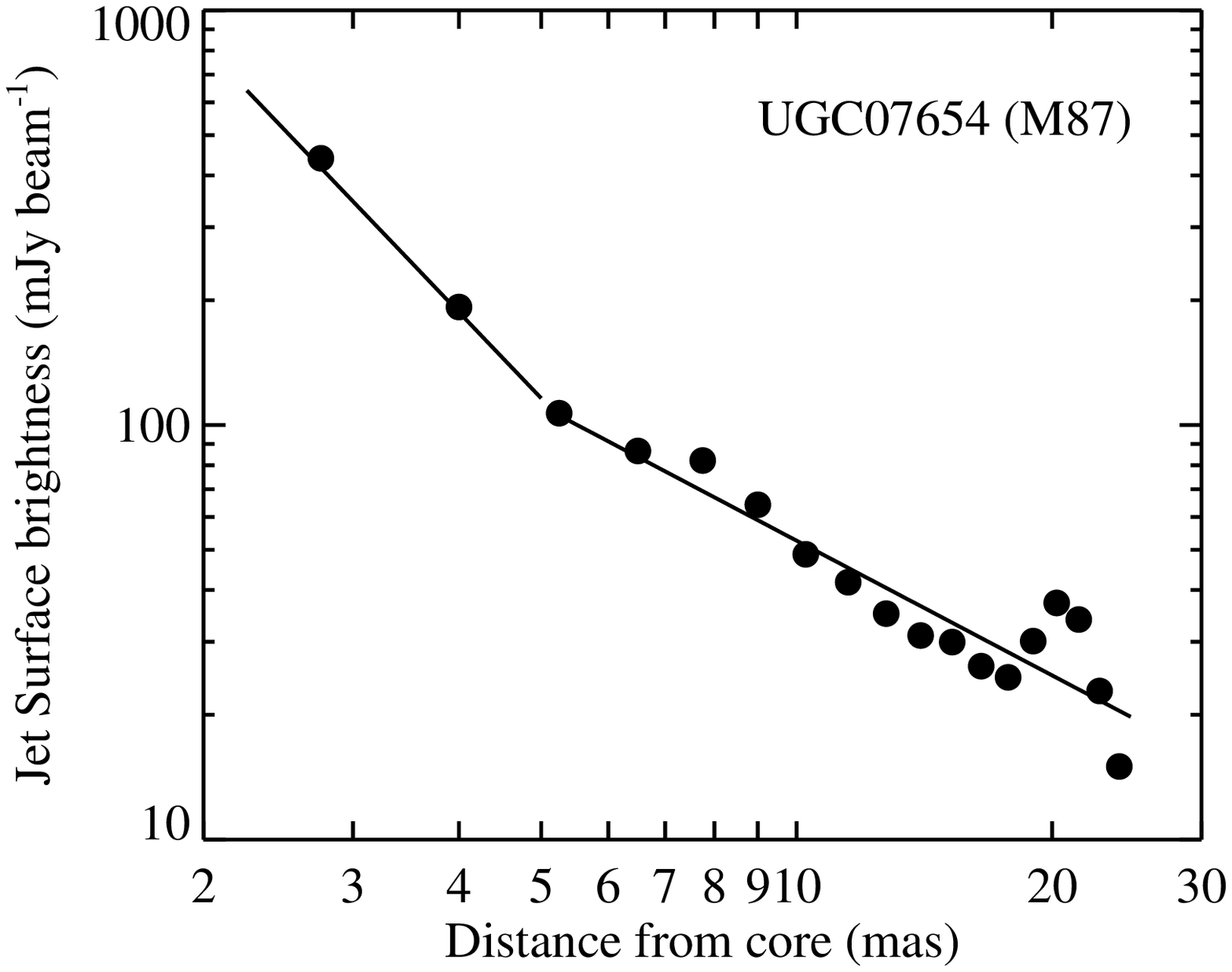}
\includegraphics[width=8.5cm]{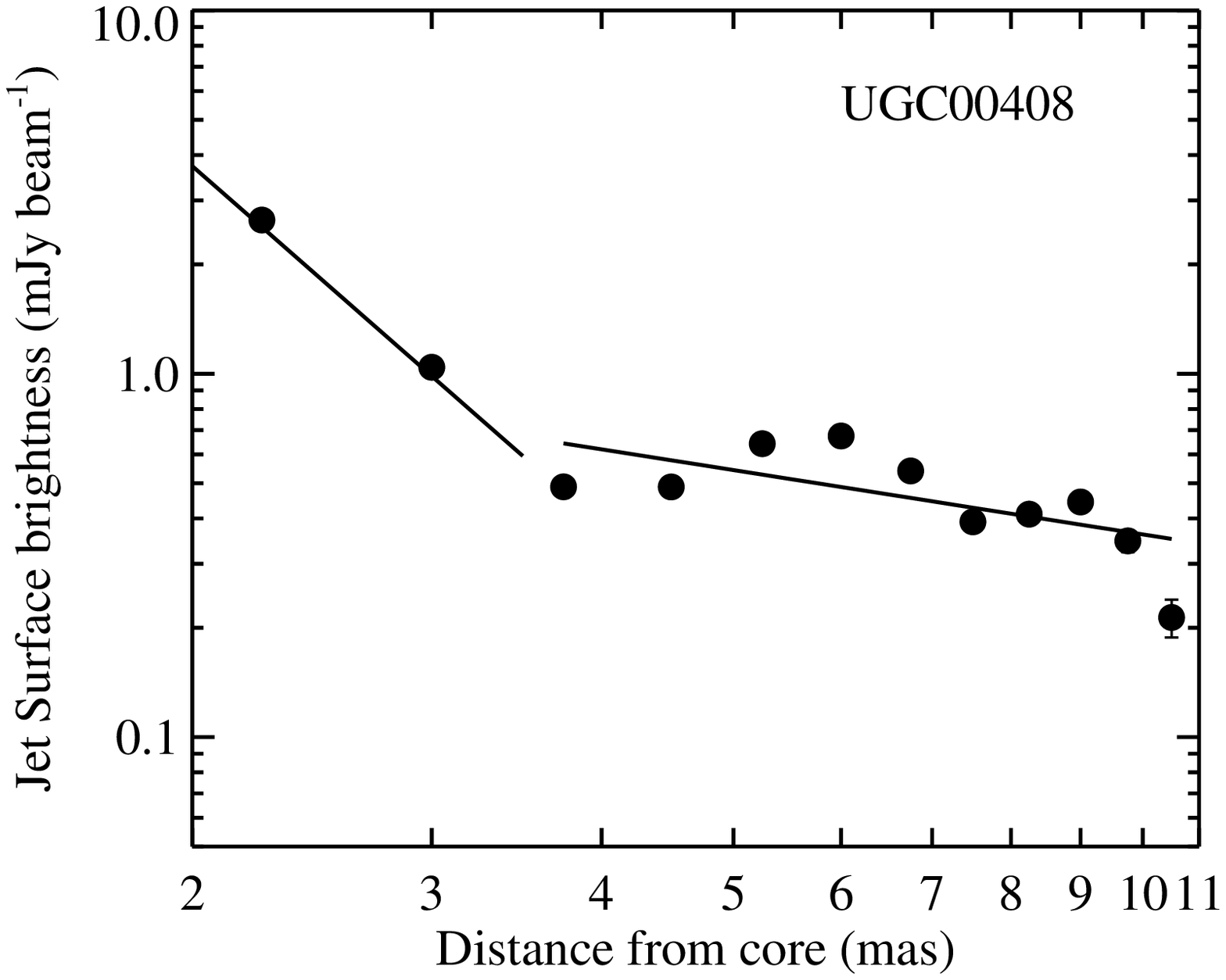}
\includegraphics[width=8.5cm]{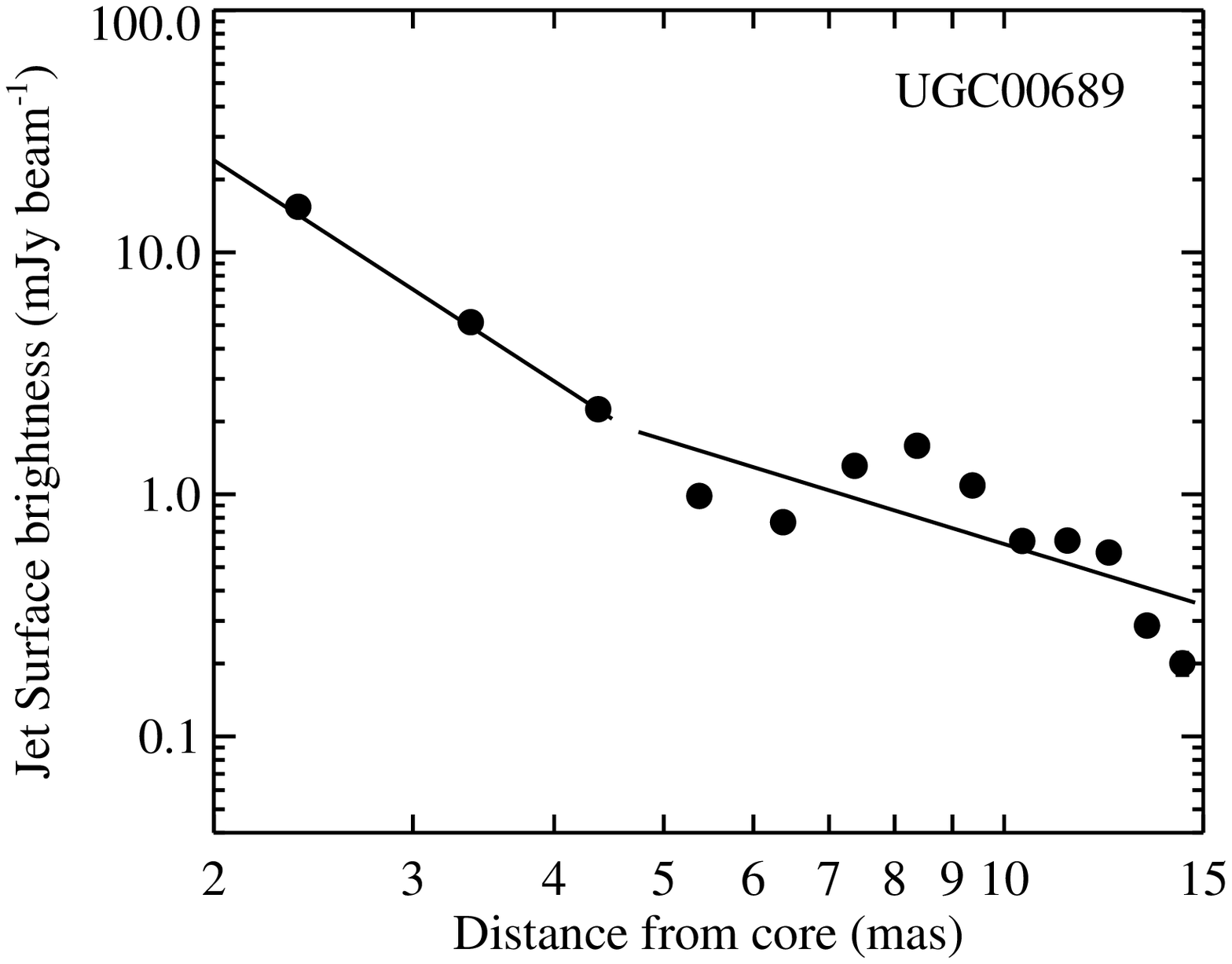}
\includegraphics[width=8.5cm]{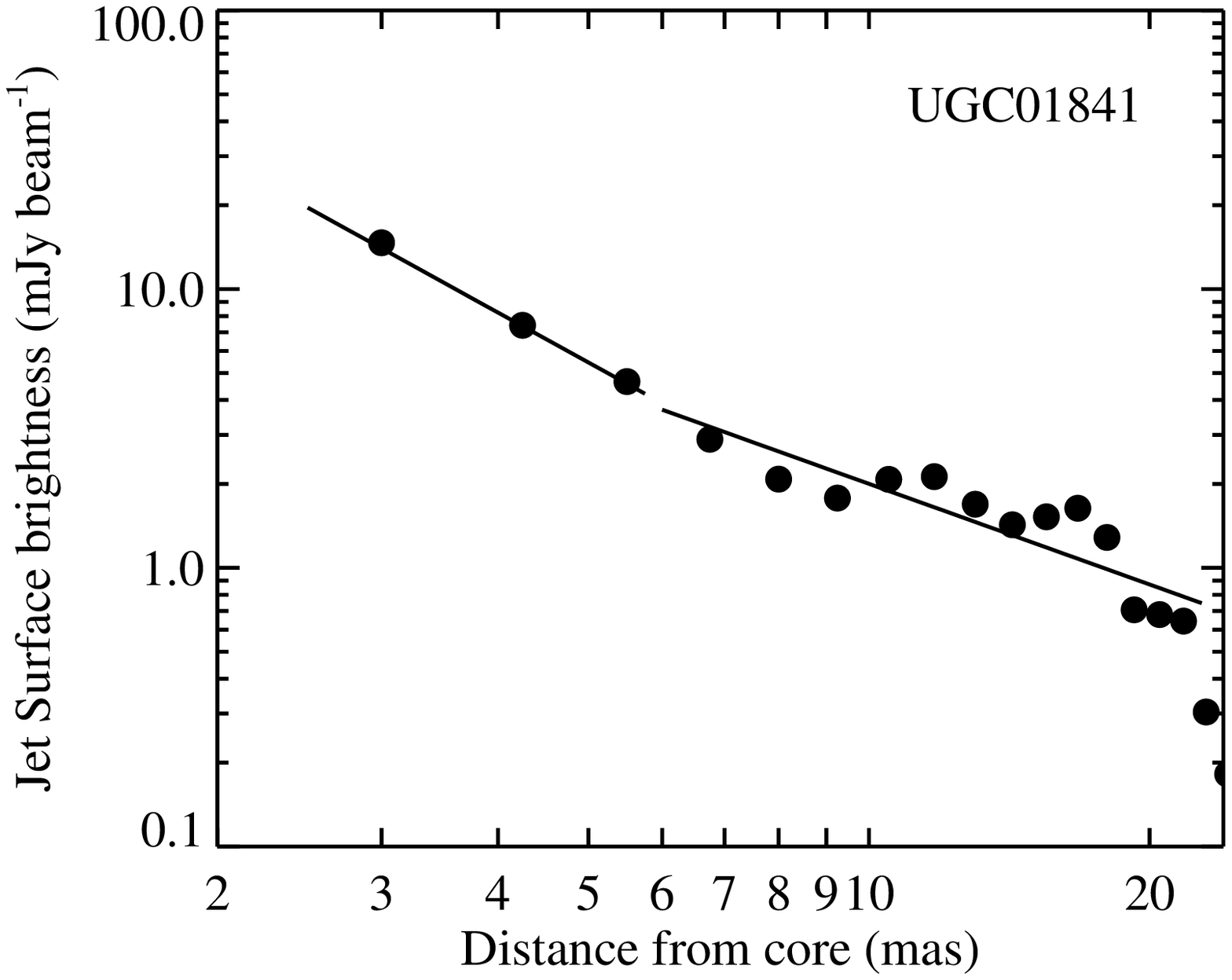}
\caption{\small Jet peak surface brightness with respect to {the} distance from the core. {\it cf.} Table \ref{jetsurface} and Section \ref{dimming} for a description of the two slopes.}
\label{figsurfa}
\end{figure}
\begin{figure}
\includegraphics[width=8.5cm]{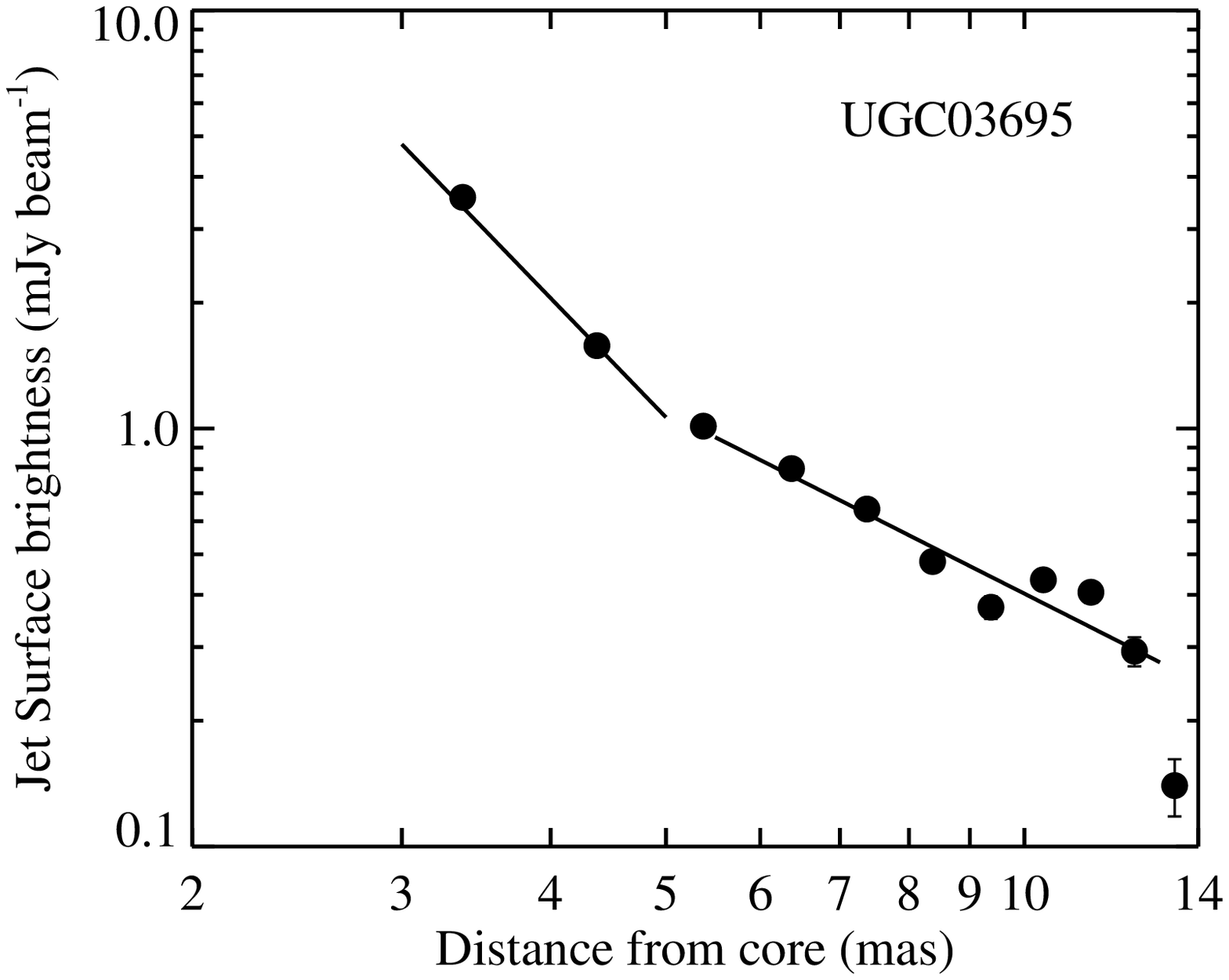}
\includegraphics[width=8.5cm]{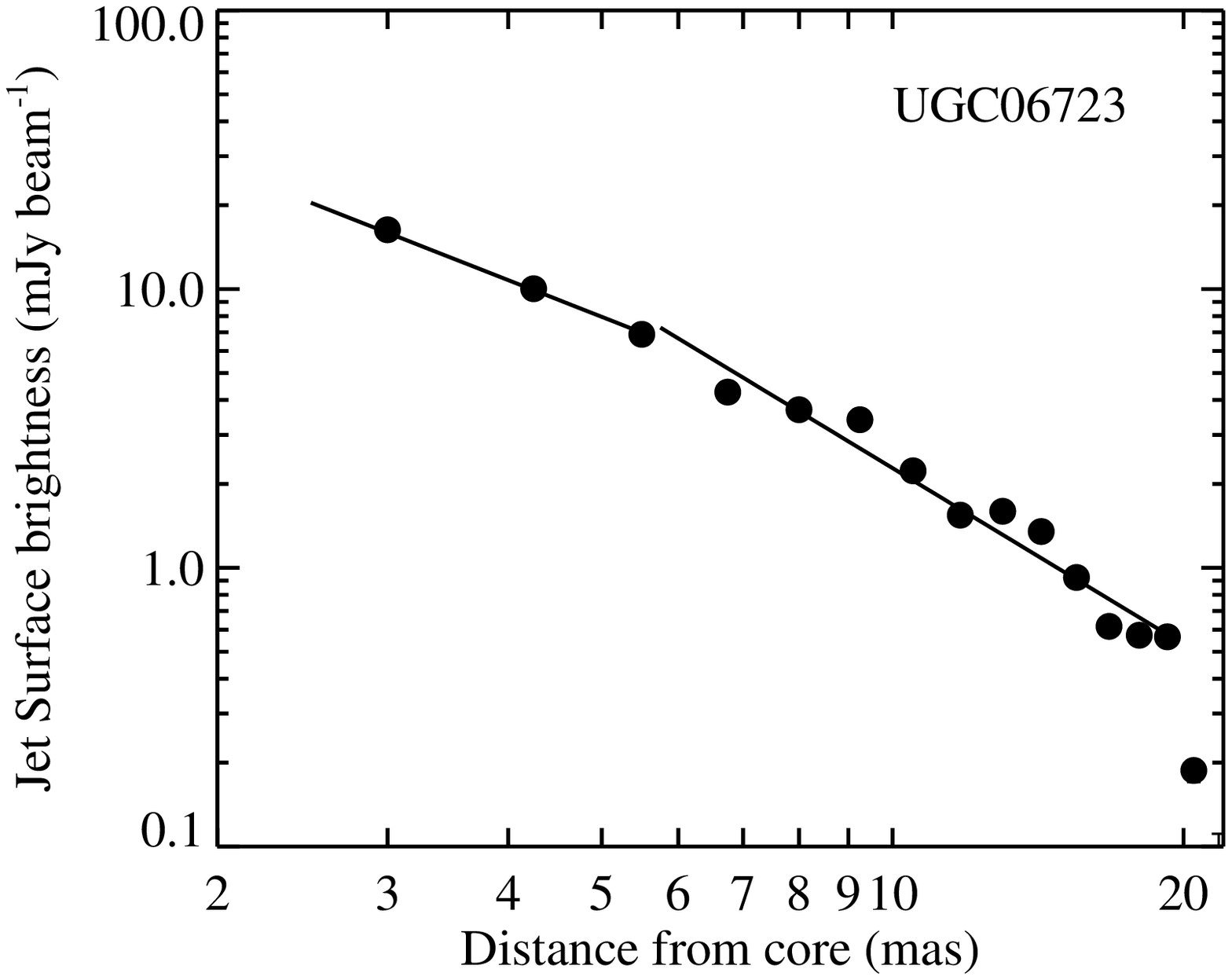}
\includegraphics[width=8.5cm]{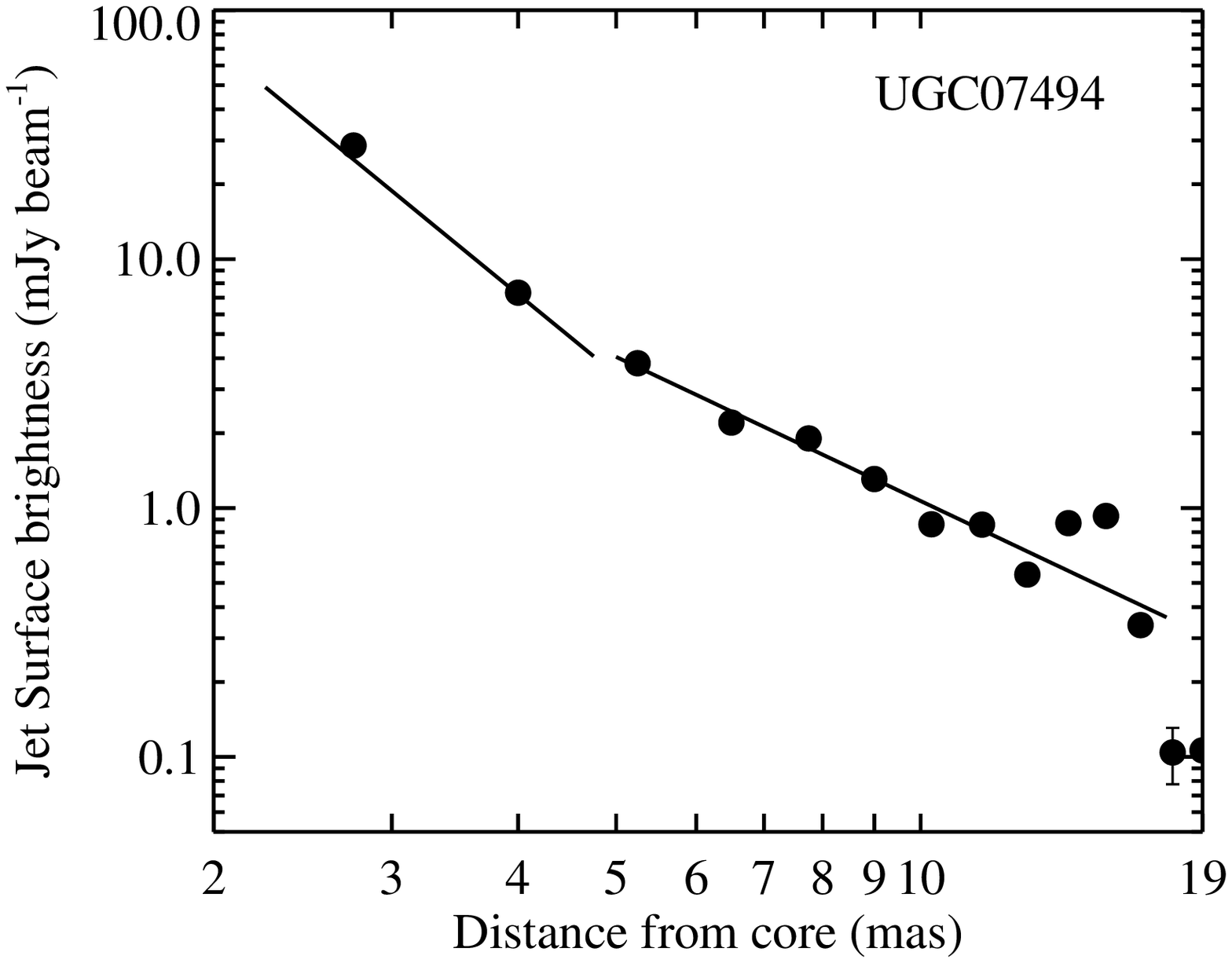}
\includegraphics[width=8.5cm]{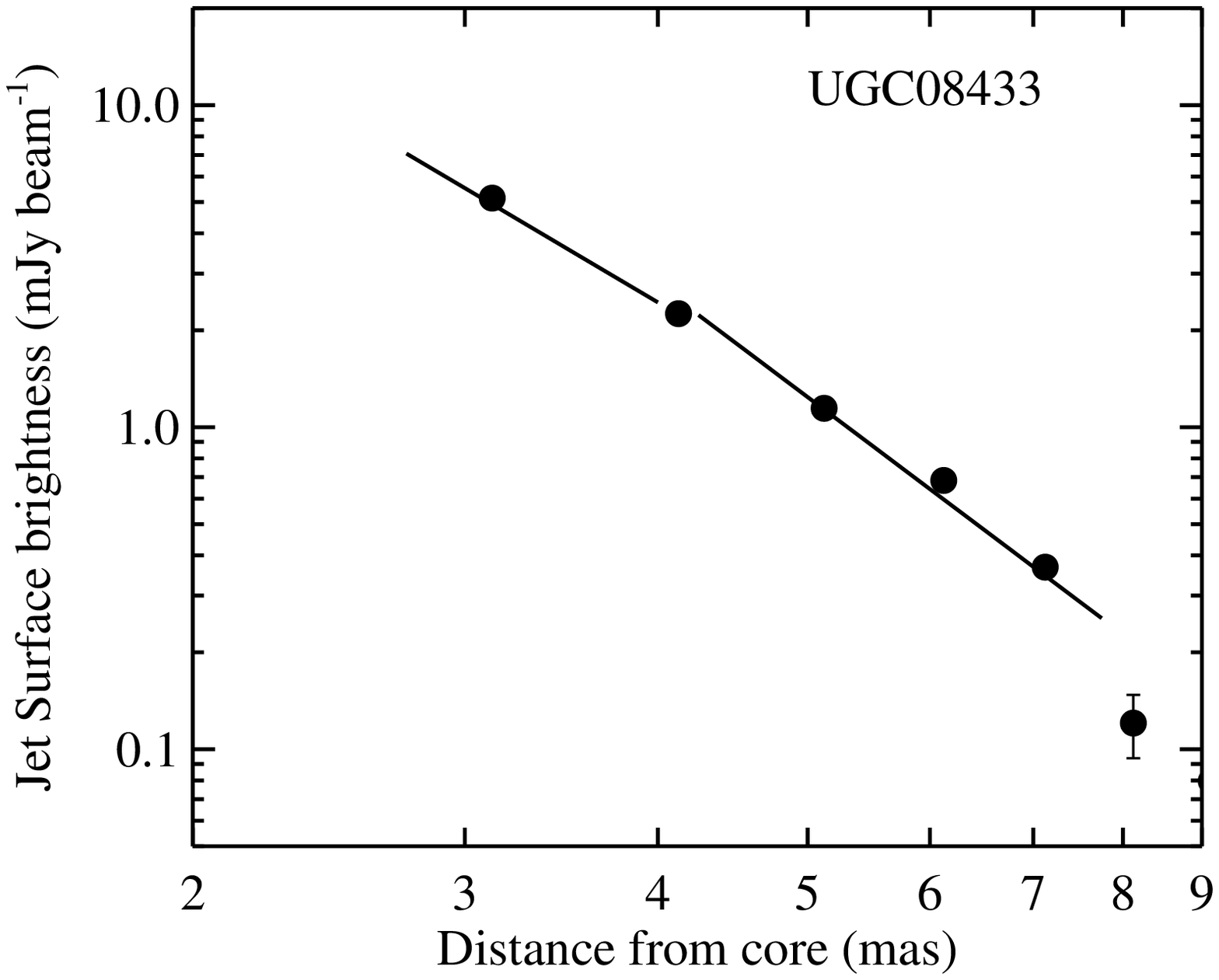}
\caption{\small Jet peak surface brightness with respect to {the} distance from the core. {\it cf.} Table \ref{jetsurface} and Section \ref{dimming} for a description of the two slopes.}
\label{figsurfb}
\end{figure}
\begin{figure}
\includegraphics[width=8.5cm]{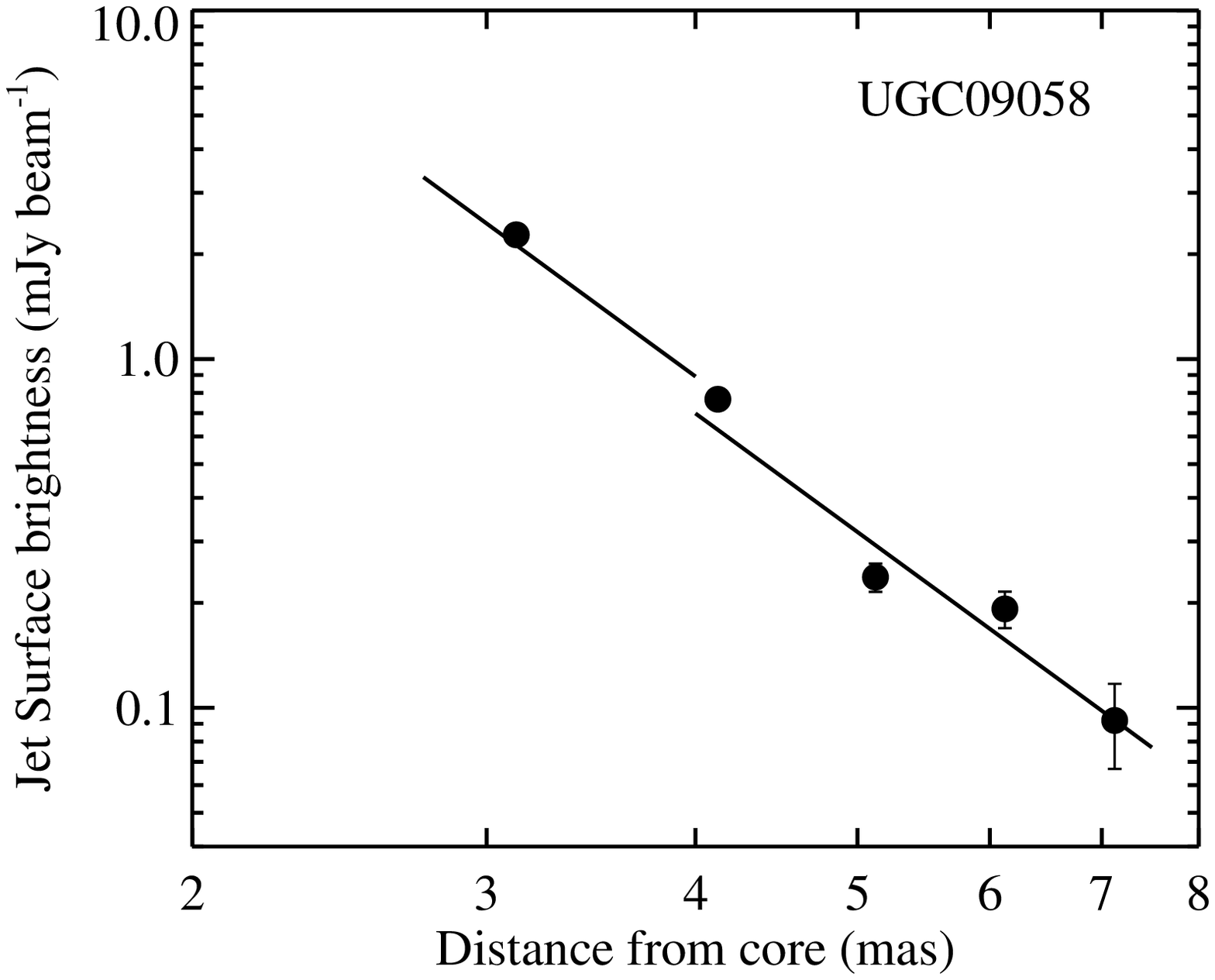}
\includegraphics[width=8.5cm]{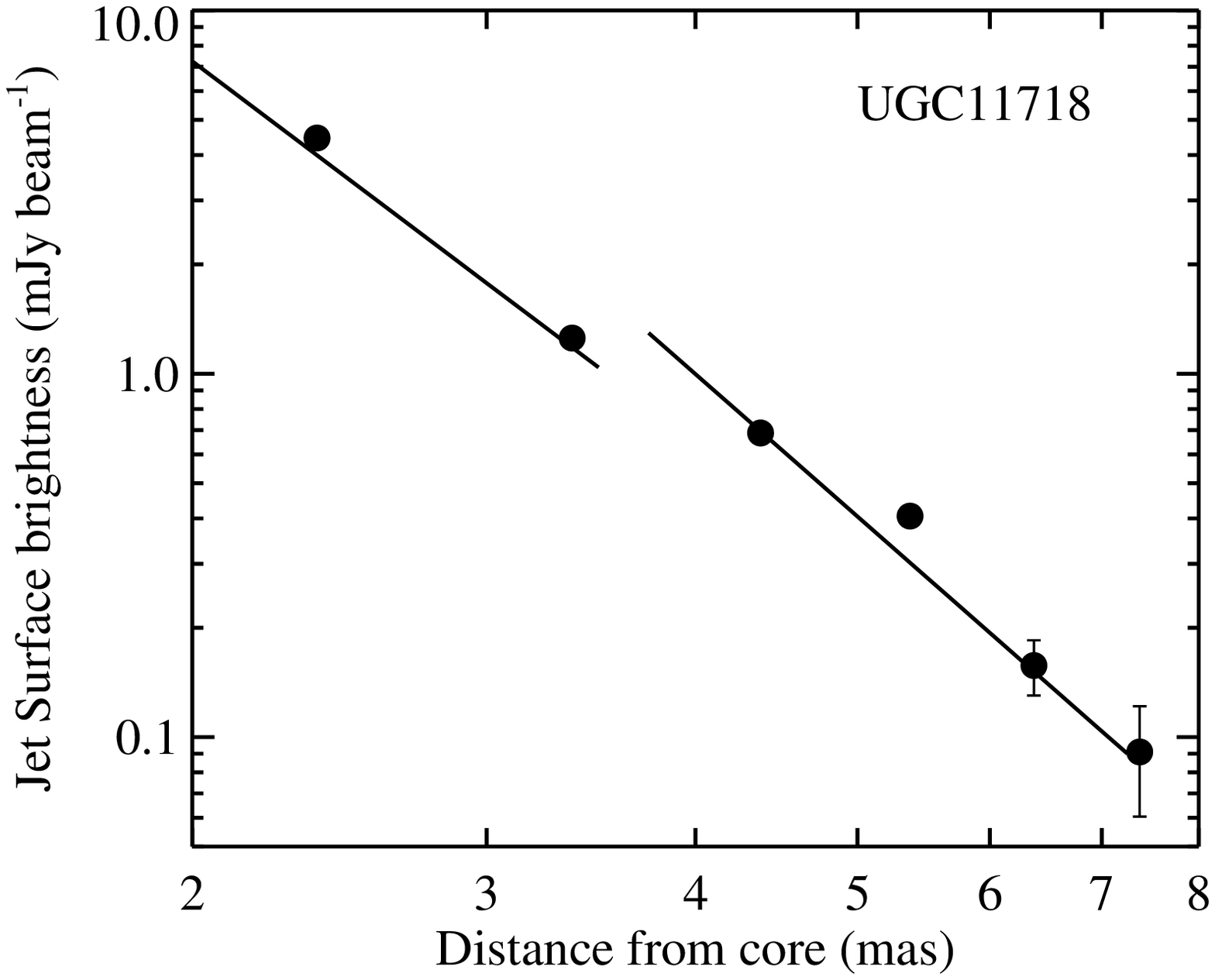}
\caption{\small Jet peak surface brightness with respect to {the} distance from the core. {\it cf.} Table \ref{jetsurface} and Section \ref{dimming} for a description of the two slopes.}
\label{figsurf}
\end{figure}

\begin{deluxetable}{ccccc}
\tabletypesize{\small}
\tablecaption{Jet Surface Brightness Evolution}
\tablewidth{0pt}
\tablehead{
\colhead{Source} &\colhead{Slope1} &\colhead{Slope1}& \colhead{Slope2} &\colhead{Break}\\
\colhead{Name}&\colhead{}&\colhead{$\#$}&\colhead{}&\colhead{mas}}
\startdata
UGC\,00408&$-$3.28$\pm$0.09& 6  &$-$0.59$\pm$0.15 &3.5\\ 
UGC\,00689&$-$3.03$\pm$0.04& 12&$-$1.43$\pm$0.22 &4.0\\ 
UGC\,01841&$-$1.85$\pm$0.04& 15&$-$1.20$\pm$0.08 &5.0\\ 
UGC\,03695&$-$2.94$\pm$0.15& 8 &$-$1.44$\pm$0.10 &5.0\\ 
UGC\,06723&$-$1.36$\pm$0.06& 15&$-$2.09$\pm$0.06 &5.5\\ 
UGC\,07494&$-$3.33$\pm$0.08& 10&$-$1.92$\pm$0.15 &5.0\\ 
UGC\,07654&$-$2.14$\pm$0.05& 10&$-$1.08$\pm$0.05 &5.0\\
UGC\,08433&$-$2.84$\pm$0.04& 4&$-$3.60$\pm$0.20 &4.0\\ 
UGC\,09058&$-$3.51$\pm$0.17& 4&$-$3.50$\pm$0.30 &4.0\\ 
UGC\,11718&$-$3.47$\pm$0.12& 8&$-$4.04$\pm$0.33 &3.5\\ 
\enddata
\tablecomments{Slope1 is derived for distances $\lesssim$5~mas from the radio core, while Slope2 is derived for distances $>$5~mas and until the jet surface brightness falls below 3 times the {\it rms} noise level. {Col.\,3: the number of data points used in the estimation of Slope1. Col.\,5: the distance from the core where the slope transitions from 1 to 2.}}
\label{jetsurface}
\end{deluxetable}

\subsection{Jet Surface Brightness ``Dimming"}
\label{dimming}
Assuming that the VLBI jet is expanding adiabatically, and the number of relativistic particles and magnetic flux is conserved, then a simple relation exists between the jet surface brightness $I_{\nu}$, jet radius $r$, and jet velocity $v$ \citep{Baum97}:
\begin{equation}
I_{\nu}\propto(\Gamma v)^{-(\gamma+2)/3}r^{-(5\gamma+4)/3}D^{2+\alpha}
\end{equation}
for a predominantly longitudinal magnetic field, and
\begin{equation}
I_{\nu}\propto(\Gamma v)^{-(5\gamma+7)/6}r^{-(7\gamma+5)/6}D^{2+\alpha}
\end{equation}
for a predominantly transverse magnetic field, where $\alpha$ is the jet spectral index, $\gamma=2\alpha$+1, $\Gamma$ is the Lorentz factor ($\equiv(1-\beta^{2})^{-1/2}$), $D$ is the Doppler factor ($\equiv(\Gamma(1-\beta cos \theta))^{-1}$), {$\beta\equiv$v/c and $\theta$ is the jet orientation angle with respect to the line of sight}. 

{Two major assumptions have been made here. The first is that there is no velocity gradient transverse to the jet, and the second is that the emission is isotropic in the jet rest frame. Taking into account the anisotropy in emission, which is important given the nature of synchrotron emission, the relations (1) and (2) above are modified to the following:
\begin{equation}
I_{\nu}\propto(\Gamma v)^{-(\gamma+2)/3}r^{-(5\gamma+4)/3}D^{3+2\alpha}
\end{equation}
for a predominantly longitudinal magnetic field, and
\begin{equation}
I_{\nu}\propto(\Gamma v)^{-(5\gamma+7)/6}r^{-(7\gamma+5)/6}D^{2+\alpha}f(D,\theta)
\end{equation}
for a predominantly transverse magnetic field \citep{Begelman93,Bondi00}. The function ``$f(D,\theta)$'' in equation 4 depends on the precise form of the field structure and must be computed numerically \citep[e.g.,][]{Laing02} for a given source. However, the anisotropy effect is much less extreme for fields without a longitudinal component and/or fields that are not highly ordered. A toroidal field, for example, would give a similar variation as the two-dimensional field sheet case, which is much less extreme (Robert Laing, private communication). 

Using therefore equations (3) and (2) for poloidal and toroidal magnetic field dominated regions, respectively, we have for $\alpha=0.7$,
\begin{equation}
I_{\nu}\propto(\Gamma v)^{-1.46}r^{-5.33}D^{4.4}
\end{equation}
\begin{equation}
I_{\nu}\propto(\Gamma v)^{-3.16}r^{-3.63}D^{2.7}
\end{equation}
}

In order to look for jet expansion on parsec-scales, we examined the deconvolved jet width $\sigma$, which is $=(\sigma_{obs}^{2} - \sigma_{beam}^{2})^{1/2}$, $\sigma_{obs}$ and $\sigma_{beam}$ being the Full Width Half Maximum (FWHM) of the observed jet width and telescope beam, respectively. For this analysis, we used images which were restored with circular beam-sizes of $2.5\times2.5$ mas, and were rotated so that the core-jet structure lay along the East-West axis (specifically at a P.A. of $-90\degr$). We found that the deconvolved jet width was typically of the order of 1 mas, which was less than half the beam-size, basically suggesting that the jets were unresolved on these scales. More sensitive, higher frequency (e.g., 15 GHz) VLBA observations are therefore required to look for jet expansion on these scales.

The peak surface brightness $I_\nu$ in the jets falls with distance $d$ from the core following a powerlaw relation, $I_\nu\propto d^{a}$ \citep[e.g.,][]{Walker87}. We examined this relation in the ten sources with jets clearly resolved along their lengths (see Figures \ref{figsurfa}$-$\ref{figsurf}). All images were first restored with circular beam-sizes of $2.5\times2.5$ mas and rotated as discussed above. Gaussians were fit to slices transverse to the jet direction using the IDL routine GAUSSFIT. The input errors on the surface brightness values were taken to be the {\it rms} noise in the radio images (we assumed a conservative noise estimate of 0.1 mJy beam$^{-1}$ for all sources). We found that for distances $\lesssim$5~mas, the peak surface brightness exhibited a steep slope with $a$ typically around $-3$ (Table \ref{jetsurface}). {We note that while the slopes were derived from over-sampled data, only the significant points spaced at about half the beam-size (i.e., at every 1.0$-$1.5 mas), have been plotted in Figures \ref{figsurfa}$-$\ref{figsurf}.} This {inner} region is likely to be affected by the presence of the bright unresolved core and jet. A slope of $-3$ in the log($I_\nu$) vs. log($d$) plot implies a Gaussian with a standard deviation of $\sim$2.3 mas, centered on the radio core peak position. 

For distances between 5 and 20 mas from the radio cores, the slope $a$ varies typically between $-1$ and $-2$ (Table \ref{jetsurface}). The average $a$ value for all ten jets turns out to be $-2.1$. However, three jets, {\it viz.,} UGC\,08433 (NGC\,5141), UGC\,09058 (NGC\,5490) and UGC\,11718 (NGC\,7052), have much steeper slopes of around $-3.5$. {This makes them shorter compared to other jets.} Excluding these three jets results in an average $a$ value of about $-1.5$. It is interesting to note that UGC\,06723 (3C\,264), UGC\,08433, UGC\,09058 and UGC\,11718 have steeper slopes for distances greater than 5 mas than for distances less than 5 mas, unlike the rest of the sources. The surface brightness slope for the jet in UGC\,09058 remains almost constant for all distances. Of these, 3C\,264 is a well-known head-tail source in a galaxy cluster and appears to be strongly interacting with the medium. NGC\,7052 on the other hand, seems to reside in a relatively isolated environment. {Therefore, it is difficult to identify any obvious reason for the steep slopes in these sources, without having more information on their jet orientations.}

Radiative losses through synchrotron and IC/CMB emission are not significant on parsec-scales as the typical electron lifetimes are of the order of a 1000 years (see Table \ref{equip}). However, inverse Compton losses from seed photons in the relativistic portion of the jets (e.g., the jet ``spine''), could still play a role in the surface brightness ``dimming''. Such a scenario, wherein the slower moving jet ``sheath'' sees the beamed radiation produced by the jet ``spine'', and undergoes inverse Compton losses, has been put forth for gamma-ray emitting AGN \citep{Ghisellini05}. Also, external photons from the accretion disk could cause inverse Compton losses in the jet electrons. 

Assuming for now that adiabatic expansion losses are primarily responsible for the jet intensity ``dimming'', two limiting cases could be considered: [1] constant jet velocity, and [2] constant jet {radius}.
\begin{enumerate}
\item
If the jet is moving at a constant velocity on parsec scales, then in order to obtain a log($I_\nu$)$-$log($d$) slope of $-1.5$, the radius of the jet $r$ should be related to distance along the jet $d$ as $r \propto d^{p}$, with {exponent} $p$ = 0.28 and 0.41 for the poloidal and toroidal magnetic fields, respectively. This implies a small jet expansion rate, which is unobservable in the 5\,GHz observations. Thus, higher frequency observations, which can resolve the jet in the transverse direction, are needed to observe this expansion. For the three sources with slopes $\sim-3.5$, $p$ would have to be 0.66 and 0.96 for the poloidal and toroidal magnetic field case, respectively. 

\citet{Walker87} have found that the jet in the radio galaxy 3C\,120 has a log($I_\nu$)$-$ log($d$) slope of $\approx-1.3$ over parsec {\it as well as on kiloparsec scales}. A much flatter slope of $\sim-0.75$ is observed in the MERLIN\footnote{Multi-Element Radio Linked Interferometer Network} jet (100 parsec scale) of 3C\,264 \citep[see][]{Baum97}. The exponent $p$ would therefore need to be 0.24 and 0.36 for 3C\,120, and 0.14 and 0.21 for 3C\,264, for poloidal and toroidal magnetic fields, respectively. Interestingly, whenever the jets are laterally resolved in 3C\,120, a $p$ value of 0.25 $-$ 0.35 is actually observed \citep[see Figure 16 of ][]{Walker87}. This is also true for the 100 parsec-scale jet of 3C\,264, where a $p$ value of 0.22 is observed for jet distances between 100 to 300 pc, and a $p$ of 0.16 for jet distances between 400 to 600 pc \citep[see Figure 10 of ][]{Baum97}. 

{\citet{Giroletti04} have reported that the jet in the BL Lac object, Mrk\,501, which could be considered a beamed counterpart of an FRI radio galaxy in the Unified Scheme, exhibits a log($I_\nu$)$-$log($d$) slope of $-1.5$ for distances $<35$ mas (=23 pc) from the core, and $-0.4$ for distances between 40 and 110 mas (=26$-$73 pc). However, unlike the case of 3C\,120, where the slope remains constant from parsec to kiloparsec scales, the slope in Mrk\,501 varies from $-1.2$, $-1.9$, $+0.6$ to $-0.9$, as the distance from the core increases from 100 to 500 mas \citep[=66$-$331 pc; see Figure 5 of ][]{Giroletti08}. For Mrk\,501, the (log of the) jet radius with respect to the (log of the) core distance has a slope of $0.5$ for distances $<35$ mas from the core, and $0.3$ for distances between 40 and 110 mas. Therefore, in sources where the jets can be resolved in the transverse direction, a slow jet expansion is indeed observed, as predicted by scenario [1].} {This makes scenario [1] a} viable option for explaining the jet intensity behavior.

\item If the jet has a constant radius on parsec scales (a cylindrical jet), then a log($I_\nu$)$-$ log($d$) slope of $-1.5$ requires that the jet speed $v$ vary with distance $d$ as $v \propto d^{q}$ (an accelerating jet!), with {exponent $q$ = 0.22, 0.15, 0.10, and 0.04,} for initial jet velocities $\beta$ = 0.6, 0.7, 0.8 and 0.9 \citep[see][]{Venturi95,Lara97}, respectively, and for distances 0.5$-$3.5 parsec from the core (typical distances covered in Figures \ref{figsurfa}$-$\ref{figsurf}), {irrespective of the dominant magnetic field structure}. Accelerating parsec-scale jets have significant implications: they are consistent with the phenomenon of ``magnetic driving'' in Poynting flux dominated jets \citep{Komissarov99,Vlahakis04}. It is worth noting that several VLBI monitoring studies have detected ``accelerating'' jet knots \citep[e.g.,][]{Hough96,Cotton99,Sudou00}, consistent with scenario [2]. 

The $q$ estimates derived above are in essence {\it upper limits}, as increasing them makes $\beta>1$, for the above assumed maximum distance and initial jet speeds. Correspondingly, {\it lower limits} to the jet inclination angles {with respect to} the line of sight, $\theta$, can be derived. For sources with log($I_\nu$)$-$log($d$) slopes of $-1.5$,  lower limits on $\theta$ must lie between 15$\degr$ and 25$\degr$ for the above jet speeds for a toroidal magnetic field dominated region; and between 30$\degr$ to 50$\degr$ for a poloidal magnetic field region. For the three sources with slopes $\sim-3.5$, lower limits on $\theta$ would need to be around 80$\degr$ to 90$\degr$ for the toroidal magnetic field region, but $>90\degr$ for the poloidal magnetic field regions, which is physically implausible. Therefore, the above scenario can explain the {entire range of} observed slopes best for the case of toroidal magnetic fields. In addition, the sources with steep slopes must lie close to the plane of the sky, consistent with the Doppler dimming effect causing a sharper drop in their jet brightness.

However, there are several cons to this model: if the jet continued to accelerate, its speed would exceed the speed of light beyond the range of distances considered here (e.g., for $d\gtrsim5$ pc). This is unphysical and would require that the jets only accelerate over the range of scales seen here, {\it i.e.,} $\sim$0.5 to $\sim$3.5 pc.
While a different range of $q$ parameters could produce the observed slopes for larger distances (say, 100 parsec scale), the $q$ values become progressively small, indicating a weaker dependence between the jet speed and distance. In addition, {the above} $q$ values depend strongly on the jet starting distance, and the total distance considered, making this case  not so well constrained. It is worth noting that VLBI monitoring also reveals jet knots that are either stationary, decelerating or moving at constant speeds \citep[e.g.,][]{Jorstad05,Homan09}. Finally, the assumption that the jet radius remains the same on kiloparsec-scales as on parsec-scales in FRI radio galaxies, is unrealistic.

{Recent relativistic numerical MHD simulations have provided invaluable insights into the collimation and acceleration of AGN jets. These simulations show that when the jet becomes unconfined and conical, the magnetic pressure gradient which accelerates the jet, is balanced by the tension of the toroidal magnetic field which slows down the jet. Therefore, jet acceleration saturates when the jet becomes conical \citep{Tchekhovskoy08,Lyubarsky09,Tchekhovskoy09}. Probing the jet in M\,87 from the sub-parsec to 100-parsec scale, \citet{Asada12} have found that the jet streamlines appear to change from parabolic to conical at a deprojected jet distance of about 250 pc. Therefore, if the FRI jets become unconfined and conical on scales of a few ten to a few hundred parsec, then jet acceleration as outlined in scenario [2] above, cannot be ruled out in {the present} data. Higher resolution polarization-sensitive radio observations which can laterally resolve the jets and indicate the magnetic field structure, as well as observations that probe jet emission from parsec to kiloparsec scales, are required to gain further insight into the jet evolution in FRIs.}
\end{enumerate}

\subsection{The Ubiquity of X-ray jets in FRIs}
15 of the sample of 21 UGC radio galaxies have been observed with {\it Chandra} \citep[see][and references therein]{Tilak07}. Of these, X-ray jets are detected in nine sources ({\it viz.,} NGC\,193, NGC\,315, 3C\,31, 3C\,66B, 3C\,264, 3C\,270, M\,84, M\,87, NGC\,5141).  The high frequency of occurance of X-ray jets in this complete sample suggests that they are a signature of a ubiquitous and important process in FRI jets. The X-ray jets in FRI radio galaxies are thought to be synchrotron emission \citep[e.g.,][]{Sambruna04,Worrall09}. The short life times of the radiating electrons in the X-rays requires in situ particle reacceleration. It appears that FRI jets start out relativistically on parsec-scales but decelerate on kiloparsec scales \citep[e.g.,][]{Laing99}. In this scenario, as the jets decelerate, they dissipate the kinetic energy flux which may power the particle acceleration. Thus, in FRI sources, the X-ray jets reveal the locations of particle reacceleration {which are} likely to be due to bulk deceleration of the jets \citep[e.g.,][]{Worrall10}.

\section{Summary and Conclusions}
We have presented here the results from a VLBA observational study of 19 UGC FRI radio galaxies at 1.6 and 5\,GHz. Polarization-sensitive VLBA observations were carried out for ten ``core-jet" sources at 5\,GHz, while second-epoch 1.6\,GHz observations were carried out for nine ``core-only" sources. {\it Chandra}-ACIS images and spectral fits for two UGC sources with X-ray jets are also presented in the paper. We examine the jet surface brightness ``dimming'' on parsec scales and derive constraints on jet evolution. The main results are summarized below:
\begin{enumerate}
\item The one-sided ``core-jet" morphology in ten sources at 5\,GHz is consistent with relativistic jet velocities on parsec-scales in these FRI radio galaxies. The second-epoch 1.6\,GHz observations failed to detect any new jets in the nine ``core-only" sources. Even more sensitive observations (with {\it rms} noise lower than $5\times10^{-5}$ Jy\,beam$^{-1}$) are therefore required to detect parsec-scale jets in them.
\item Seven of the ten sources studied show some polarization in their parsec-scale jets at 5\,GHz. UGC\,06723 (3C\,264) shows the most extensive polarization in its jet, with the polarization electric vectors suggesting a complex magnetic field structure reminiscent of a ``spine-sheath'' structure. Polarization seems to be detected close to the jet edge in a couple of sources with an aligned magnetic field ($B_{||}$) geometry, which could either be suggestive of ``shearing'' due to jet-medium interaction or a helical magnetic field. In a few cases the polarization electric vectors are oblique with respect to the jet direction. This could indicate a {Faraday} screen of ionized gas that may have rotated the polarization electric vectors, or  the presence of oblique shocks. {A thin outer layer of the jet could in principle serve as the Faraday screen.}
\item The peak intensity ($I_{\nu}$) in the parsec-scale jets falls with distance ($d$) from the core following a powerlaw relation, $I_{\nu}\propto d^{p}$, with $p$ varying between $-1$ and $-2$ (typical $p\approx-1.5$). Three sources {with short jets} have steeper slopes of about $-3.5$.
\item On the assumption that adiabatic expansion losses are primarily responsible 
for the jet surface brightness ``dimming'', two limiting cases are considered: [1] the jet has a constant speed on parsec-scales and is expanding gradually, but the expansion is not visible in the 5\,GHz images, or [2] the jet has a constant radius, and is accelerating on parsec-scales. In order to explain the whole range of observed log($I_{\nu}$) vs. log($d$) slopes, the jet must be dominated by toroidal magnetic fields, in the latter case. Accelerating parsec-scale jets are consistent with the phenomenon of ``magnetic driving'' in Poynting flux dominated jets. The main caveat to case [2] is that if the jet continued to accelerate beyond the narrow range of distances considered here, its speed would become larger than the speed of light, which is clearly unphysical. {However, if the jets become conical on 100-parsec scales, as suggested in M\,87 \citep{Asada12}, then case [2] may still be viable. While slow jet expansion is indeed indicated in a few sources that have been resolved in the transverse direction, such as Mrk\,501, 3C\,264 and 3C\,120, on scales of ten to hundred parsec, in agreement with the predictions of case [1], case [2] cannot be ruled out in the present data. Radio observations that probe jet emission all the way from parsec to kiloparsec scales are required to fully understand the jet evolution in FRI radio galaxies.}   
\item The X-ray image of UGC\,00408 (NGC\,193) shows the presence of an X-ray bubble 
{with a cavity radius of $\sim$16 kpc. We estimate that the work done on the X-ray
gas by the inflation of the bubble is $\approx$7E+58 erg.}
\item  Of the 15 sample UGC radio galaxies observed with {\it Chandra}, X-ray jets are detected in nine sources. The high frequency of occurance of X-ray jets suggests that they are a signature of a ubiquitous and important process in FRI jets. It appears that the FRI jets start out relativistically on parsec-scales but decelerate on kpc scales, and the X-ray emission reveals the location of particle reacceleration due to bulk deceleration in their jets.
\end{enumerate}

\acknowledgments
We thank the referee for her/his comments which have lead to considerable improvement of this paper. We thank Robert Laing, Alexander Tchekhovskoy, Masanori Nakamura and Grant Tremblay for making valuable suggestions which have further improved this work. The National Radio Astronomy Observatory is a facility of the National Science Foundation operated under cooperative agreement by Associated Universities, Inc. This work is supported in part by the Radcliffe Institute for Advanced Study at Harvard University.

\bibliographystyle{apj}
\bibliography{ms}

\end{document}